\documentclass[fleqn,usenatbib]{mnras}
\usepackage{mathptmx}
\usepackage[varvw]{newtxmath}  %evs: Added to prevent cursive v to look like $\nu$.
\usepackage[T1]{fontenc}
\usepackage{ae,aecompl}
\usepackage{times}
\usepackage{amsmath}
\usepackage{upgreek}
\usepackage{xcolor}
\usepackage{graphicx}
\usepackage{pdflscape}
\usepackage{multicol} 
\usepackage{amsmath,bm}
\usepackage{caption}
\title{Formation of hub-filament structure triggered by cloud-cloud collision in W33 complex}
\author[J. W. Zhou]{
Jian-Wen Zhou \thanks{E-mail: a18893153117@163.com}$^{1,2}$
Shanghuo Li,$^{3}$
Hong-Li Liu,$^{4}$
Yaping Peng,$^{5}$
Siju Zhang,$^{6}$
Feng-Wei Xu,$^{6,7}$
\newauthor
Chao Zhang,$^{8}$
Tie Liu,$^{9}$
Jin-Zeng Li,$^{1}$
\\
Affiliations are listed at the end of the paper}

\date{Accepted XXX. Received YYY; in original form ZZZ}
\pubyear{2021}
\begin{document}
\label{firstpage}
\pagerange{\pageref{firstpage}--\pageref{lastpage}}
\maketitle

\begin{abstract}
Hub-filament systems are suggested to be birth cradles of high-mass stars and clusters, but the formation of hub-filament structure is still unclear. Using the survey data FUGIN $^{13}$CO (1-0), C$^{18}$O (1-0), and SEDIGISM $^{13}$CO (2-1), we investigate formation of hub-filament structure in W33 complex.
W33 complex consists of two colliding clouds, called W33-blue and W33-red.
We decompose the velocity structures in W33-blue by fitting multiple velocity components, and find a continuous and monotonic velocity field. Virial parameters of {\it Dendrogram} structures suggest the dominance of gravity in W33-blue. The strong positive correlation between velocity dispersion and column density indicates the non-thermal motions in W33-blue may originate from gravitationally driven collapse. These signatures suggest that the filamentary structures in W33-blue result from the gravitational collapse of the compressed layer. However, the large scale velocity gradient in W33-blue may mainly originate from the cloud-cloud collision and feedback of active star formation, instead of the filament-rooted longitudinal inflow. From the above observed results, we argue that the cloud-cloud collision triggers formation of hub-filament structures in W33 complex. Meanwhile, the appearance of multiple-scale hub-filament structures in W33-blue is likely an imprint of the transition from the compressed layer to a hub-filament system.
\end{abstract}


\section{Introduction} \label{introduction} 
Filamentary structures are ubiquitous in molecular clouds, and sevaral formation mechanisms of it have been advocated, such as the gravitational instability of a flattened isothermal cloud, supersonic MHD isothermal turbulence, turbulence in thermally unstable gas, global gravitational contraction and others \citep{ Miyama1987-78,Padoan2001-553,Balsara2001-327,Hartmann2007-654,Banerjee2009-398,Vazquez2009-707, Schneider2010-520,Andre2010-518,Arzoumanian2011-529,Kirk2013-766,  Andre2014, Gomez2014-791, Padoan2016-822,  Vazquez2019-490, Padoan2020-900}. 
A collapsing cloud with converging filaments forms a hub-filament system, in such systems, converging flows are funneling materials into the hub through the filaments, then part of embedded cores can grow into massive stars due to the sustained supply of materials from the filamentary environment. Therefore, hub-filament systems are suggested to be birth cradles of high-mass stars and clusters
\citep{Myers2009-700,Schneider2012-540,Peretto2013, Henshaw2014-440,Yuan2018-852,Lu2018-855,Liu2019-487,Liu22,Trevino2019-629,Dewangan2020-903,Zhou2022-514}. However, so far, the definition of hub-filament is not clear, and the formation of hub-filament structure is still under debate. As discussed in \citet{Zhou2022-514}, the prevalent hub-filament systems found in proto-clusters favors the pictures advocated by either global hierarchical collapse (GHC) or inertial-inflow scenarios \citep{Vazquez2019-490,Padoan2020}, both of them emphasize longitudinal inflow along filaments. However, the mechanism of filament formation in the two models are completely different \citep[e.g.,][for a review]{Liu22a}. The GHC model predicts anisotropic gravitational contraction with longitudinal flow along filaments at all scales, filaments represent the locus of gas being accreted from the cloud to the star-forming clumps, thus the filaments are not density but flow structures as the gas within them is constantly replenished. The inertial-inflow model emphasizes the large-scale turbulent compression, where filament is the location of the intersection of postshock layers. Gravity or turbulence, which is dominant? It’s a crucial question to distinguish the driving mechanism of filament formation between different theoretical models and understand the underlying dynamics in star formation regions \citep{Ballesteros2018-479}. This question also tightly correlates with the formation of hub-filament structure. 

Currently, some theoretical models and simulations of hub-filament formation have been proposed. \citet{Burkert2004-616} presents two-dimensional simulations of finite, self-gravitating gaseous sheets. % such configurations are subject to global collapse. 
They emphasize the long-range effects of gravity and the importance of cloud boundaries in generating structure and supersonic turbulence. % Cluster-forming gas is often concentrated as a result of gravity acting on irregular boundaries, this mechanism leads to a very rapid gas infall, and then drives the formation of massive stars. 
The simulations have interesting implications for the gravitational evolution of overall molecular cloud structure, and such clouds might originate as roughly sheetlike sections of gas accumulated as a result of large-scale flows in the local interstellar medium. Fig.12 of this work shows the results of the collapse of a sheet with uniform initial surface density but an arbitrary irregular boundary, this impressive figure presents similar morphology to W33-blue, thus perhaps reveals some intrinsically physical processes regarding the formation of hub-filament structure in W33-blue.
\citet{Myers2009-700} suggests that the hub-filament structure observed in molecular clouds may be a “fingerprint” of the compression which reshapes molecular clouds into modulated layers. In this picture, the compression of low-density gas into modulated layers can produce filaments, while hub can originate from compression of a centrally condensed clump. The innermost parts of the layer have the densest gas, the shortest free-fall time, and thus the greatest likelihood of fragmentation and star formation. Here, filaments originate from compression of external pressure, such as winds and shocks, this scenario can be reproduced in the inertial-inflow model \citep{Padoan2020-900}. However, it is also possible that the observed convergence of filamentary arms onto hubs has a more dynamic basis, such as gravitational inflow of filament gas toward the attracting hub \citep{Balsara2001-327,Banerjee2006-373,Myers2009-700}. This scenario is closer to the GHC model \citep{Vazquez2009-707,Vazquez2019-490}. 
\citet{Gomez2014-791} performed a smoothed particle hydrodynamics (SPH) simulation of the formation of a molecular cloud from a convergent flow of diffuse gas. Here, the turbulent motions are unable to stabilize the cloud and soon its dynamics becomes dominated by gravity, and the cloud begins to collapse. The mass of cloud continues to grow due to accretion, and soon it contains a large number of Jeans masses \citep{Vazquez2007-657}, naturally leading to formation of filaments. When the initially planar cloud collapses onto filaments, the filaments themselves fall into the largest-scale potential well and eventually merge into each other, forming a larger-scale hub-filament system towards the end of the simulation. Similarly, in the SPH simulation of \citet{Balfour2015-453}, hub-filament structure can form in the process of cloud-cloud collision. The shock-compressed layer can break up into an array of predominantly radial filaments or a network of filaments, depending on the collision velocity. Once this layer becomes sufficiently massive, it fragments gravitationally to produce filaments. At the same time, the shock-compressed layer contracts heterogeneously due to the non-homogeneous surface density, and consequently the inner parts of the layer converge on the center on a shorter time scale than the outer parts. This has the effect of stretching filaments into predominantly radial orientations. 
In summary, the bone of contention of these simulations and models:
are the filaments of hub-filament systems gravitational collapse structures or turbulent compression structures ? In this paper, we select molecular cloud complex (hereafter W33 complex) with a giant hub-filament structure to discuss this issue in detail.

W33 complex is a massive star-forming region located at  $l\sim 12^\circ.8$ and $b\sim -0.2^\circ$ in the Galactic plane. The parallactic distance of W33 was measured to be 2.4 kpc based on the water maser observations by \citet{Immer2013-553}. W33 harbors many star forming clumps, OB stars, and H{\sc ii} regions. Within an area of 15 pc $\times$ 15 pc, there are six dust clumps (W33 Main, W33 A, W33 B, W33 Main1, W33 A1, and W33 B1) revealed by the Atacama Pathfinder Experiment (APEX) Telescope Large Area Survey of the GALaxy (ATLASGAL) 870$\mu$m survey \citep{Schuller2009-504, Contreras2013-549, Urquhart2014-568}. Molecular line observations show a complex velocity filed toward the W33 complex.  \citep{Gardner1972-12,Goldsmith1983-265,Immer2013-553,Kohno2018-70S}. \citet{Immer2013-553} detected water maser emission at a velocity of $\sim$35 km s$^{-1}$ toward W33 Main and W33 A and $\sim$60 km s$^{-1}$ toward W33 B, finding that these two velocity components are at the same annual parallactic distance of 2.4 Kpc. \citet{Kohno2018-70S} reported three velocity components at 35, 45, and 58 km s$^{-1}$, and found the signatures of the collision between two clouds at 35 and 58 km s$^{-1}$. Therefore, a cloud-cloud collision (CCC) scenario was proposed to explain the massive star formation in W33 complex. Such study was extensively carried out in \citet{Dewangan2020-496}, they also examined the large-scale environment of W33 complex. \citet{Liu2021-646} identified a hub-filament system in the velocity range [30, 38.5] km~s$^{-1}$ by C$^{18}$O (1-0) emission, and argued that the gas flows along the tributary filaments in the system which probably promote the proto-cluster formation in W33 complex. 

In this paper, we aim to associate the formation of hub-filament structure and the cloud-cloud collision in W33 complex together, and investigate the role of compression and gravitational collapse in the formation of hub-filament system. We also examine some aforementioned models and simulations. 

% To explain star formation (SF) activity in the direction of W33 A, \citet{galvan10} proposed a triggered SF scenario by filamentary convergent gas flows from two different velocity components. \citet{messineo15} also mentioned the possibility of sequential SF and feedback in W33. Using the high-resolution molecular line data, the authors found spiral and filamentary structures around the central massive young stellar object in W33 A Maud et al. (2017). 
% HFS cannot originate by compression of a purely uniform medium, because that would only produce filaments, and no hubs. Similarly, HFS cannot originate by compression of a centrally condensed clump, because that would produce a hub but no nearly parallel filaments. Only with the combination of both clump and medium does the characteristic HFS pattern arise.

\section{Observations and data reduction}\label{sec:observations}
\subsection{Survey data}
\label{surveys}
The spectral lines $^{12}$CO (1$-$0), $^{13}$CO (1$-$0), and C$^{18}$O (1$-$0) were retrieved from the FOREST Unbiased Galactic plane Imaging survey with the Nobeyama 45-m telescope \citep[FUGIN;][]{Umemoto2017-69} survey, 
which uses the main beam temperature ($T_\mathrm{mb}$) in calibration
 \citep{Umemoto2017-69}. 
The typical root-mean-square (rms) noise level\footnote[1]{https://nro-fugin.github.io/status/} is $\sim$1.5~K, $\sim$0.7~K, and $\sim$0.7~K for $^{12}$CO, $^{13}$CO, and C$^{18}$O lines, respectively.
% The multi-beam receiver, FOur-beam REceiver System on the 45-m Telescope \citep[FOREST;][]{Minamidani2016-9914}, was utilized for the FUGIN survey. 
FUGIN is the first CO Galactic plane survey at 20$\arcsec$ resolution in the $^{12}$CO, $^{13}$CO, and C$^{18}$O $J=1-0$ lines simultaneously. The effective angular resolution of the final map is 20$\arcsec$ for $^{12}$CO, 21$\arcsec$ for $^{13}$CO and C$^{18}$O. Maps are produced as a single FITS cube by gridding baseline-subtracted and scaled data over the desired region. The Bessel $\times$ Gaussian function is used as the convolution function, and data are gridded to 8.5$\arcsec$ and 0.65 km~s$^{-1}$ for the 1st quadrant 
(10$^{\circ} \leq l \leq 50^{\circ}$, $\left|b\right| \leq 1^{\circ}$; 80 deg$^{2}$ ). Considering the adopted window function, effective velocity resolution is 1.3 km~s$^{-1}$ at 115 GHz. 

%FUGIN is the first CO Galactic plane survey at $\timeform{20''}$ resolution in the $^{12}$CO, $^{13}$CO, and C$^{18}$O $J=1-0$ lines simultaneously. The effective angular resolution of the final map is $\timeform{20''}$ for $^{12}$CO, $\timeform{21''}$ for $^{13}$CO and C$^{18}$O. Maps are produced as a single FITS cube by gridding baseline-subtracted and scaled data over the desired region. The Bessel $\times$ Gaussian function is used as the convolution function, and data are gridded to $\timeform{8''.5}$ and 0.65 km~s$^{-1}$ for the 1st quadrant ($\timeform{10^{\rm o}} \leq l \leq \timeform{50^{\rm o}}$, $|b| \leq \timeform{1^{\rm o}}$; 80 deg$^{2}$). Considering the adopted window function, effective velocity resolution is 1.3 km~s$^{-1}$ at 115 GHz. 

The SEDIGISM survey \citep{Schuller2017-601} covers the inner Galactic plane between $-60^{\rm o} \leq \ell \leq 18^{\rm o}$ and $|\textit{b}|\leq 0.5^{\rm o}$, which was observed from 2013 to 2016 with the SHeFI heterodyne receiver \citep{Vassilev2008-490} at the APEX telescope. The prime targets of the survey are the $^{13}$CO (2-1) and C$^{18}$O (2-1) molecular lines. The average rms noise of the survey is $\rm 0.9~K$ (T$_{\rm MB}$) at a velocity resolution of $\rm 0.25~km\,s^{-1}$, and a full width half maximum (FWHM) beam size of $30\arcsec$, and a pixel-size of $9.5\arcsec$ \citep{Schuller2021-500}. 
% The velocity resolution of SEDIGISM survey is much better than FUGIN survey, thus we combine these two datasets together in this paper.

\subsection{Archived data}
Auxiliary data of images were complemented by the GLIMPSE \citep{Benjamin2003} and Herschel Infrared GALactic plane survey \citep{Molinari2010-518,Molinari2016-591}. The images of the Spitzer Infrared Array Camera (IRAC) at 8.0 $\mu$m were retrieved from the \emph{Spitzer} Archive, the angular resolutions of the images in the IRAC bands are $< 2\arcsec$. 
All of the Hi-GAL data have been processed by PPMAP procedure, by which high resolution column density and dust temperature maps could be obtained. The resolution of column density and dust temperature maps is $\sim12\arcsec$, and the resulting data are available online\footnote{http://www.astro.cardiff.ac.uk/research/ViaLactea/} \citep{Marsh2016-459,Marsh2017-471}.

\section{Results and data analysis} \label{sec:analysis}

\subsection{Hub-filament structure of W33-blue}\label{w33-blue}

\begin{figure*}
%   \centering
  \includegraphics[width=1\textwidth]{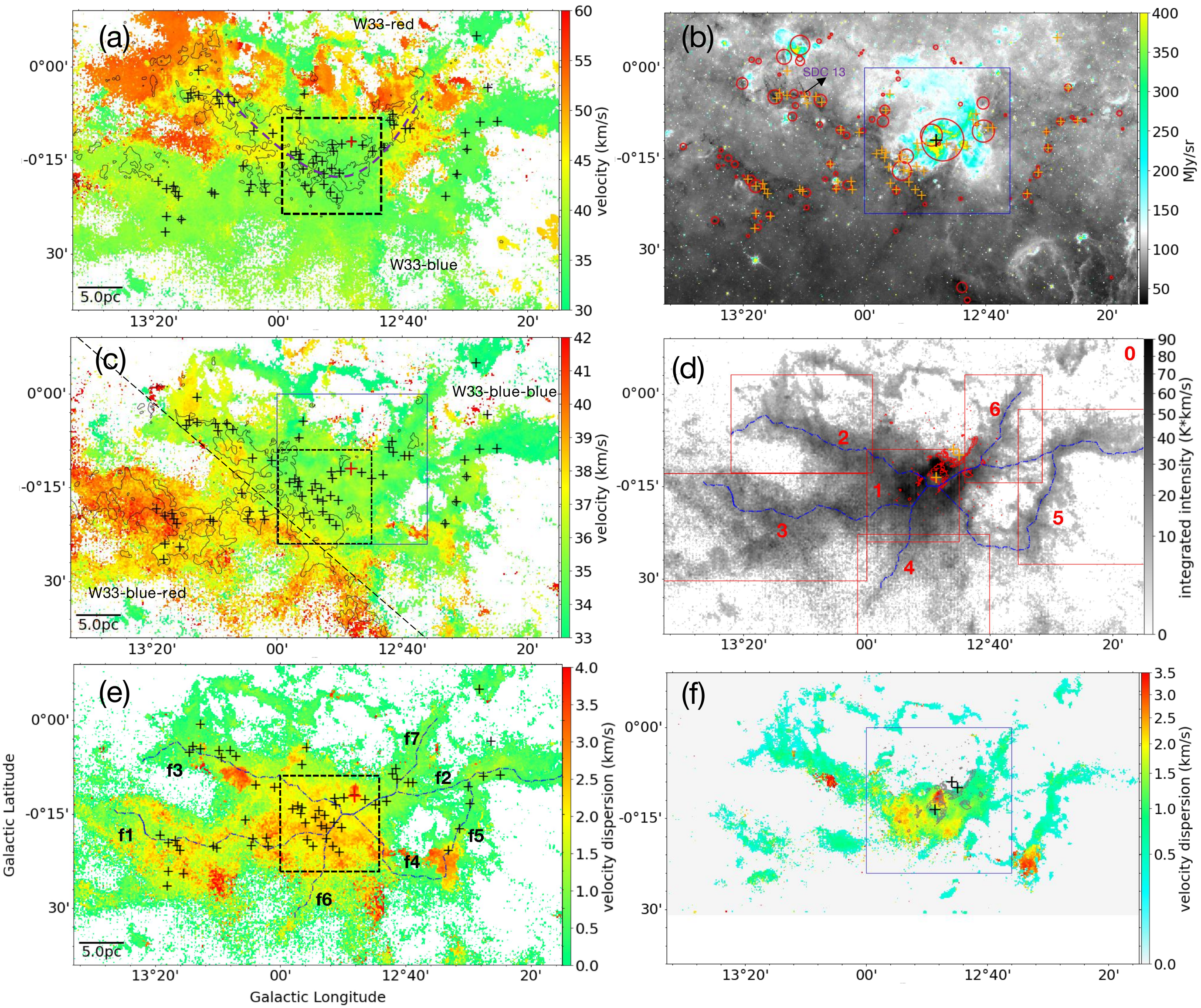}
\caption{
(a) Moment-1 map of $^{13}$CO (1$-$0) for the entire W33 complex, made over the velocity range [29.6, 60.2] km s$^{-1}$. The contour is the \emph{Herschel} H$_{2}$ column density with a level of 3.35$\times$10$^{22}$ cm$^{-2}$, dashed curve marks the bent compression structure due to cloud-cloud collision. Red “+” represents the position of W33-main, while black “+” marks the ATLASGAL clump. Dashed black box marks the bottom of collision crater, where has the most concentrated clump distribution; 
(b) The colour images of \emph{Spitzer} 8 $\mu$m emission, black “+” represents the position of W33-main, orange “+” marks the ATLASGAL clump. The red circle shows the IRDC, where the size of the circle reflects the radius of IRDC; 
(c) Moment-1 map of $^{13}$CO (1$-$0) for W33-blue over the velocity range [29.6, 43.3] km s$^{-1}$. The contour is the \emph{Herschel} dust temperature with a level of 18.6 K, black “+” marks the ATLASGAL clump. Dashed straight line roughly displays the boundary of two velocity components in W33-blue; 
(d) The integrated intensity map of $^{13}$CO (1$-$0), velocity range [29.6, 43.3] km s$^{-1}$. Blue dashed lines display the filament skeletons identified by FILFINDER algorithm according to the integrated intensity map. Red contours show the peak emission of \emph{Spitzer} 8 $\mu$m. Three orange “+” represent O4–7(super)-giant stars, red boxes mark several sub-regions from “1” to “6”, “0” represents the entire region.
(e) The velocity dispersion of W33-blue derived from the moment-2 map of $^{13}$CO (1$-$0) in the velocity range [29.6, 43.3] km s$^{-1}$, black “+” marks the ATLASGAL clump, blue dashed lines display the filament skeletons. Dashed black box marks the bottom of collision crater; 
(f) The velocity dispersion of W33-blue derived from the moment-2 map of $^{13}$CO (2$-$1), gray contours show the peak emission of \emph{Spitzer} 8 $\mu$m,  and the black “+” represents O4–7(super)-giant stars. Blue boxes in figures (b), (c) and (f) show the region investigated by \citet{Kohno2018-70S}.
}
\label{vf}
\end{figure*}

In Fig.5, Fig.6, and Fig.7 of \citet{Umemoto2017-69}, clear hub-filament structure can be traced by $^{13}$CO (1$-$0) compared with $^{12}$CO (1$-$0) and C$^{18}$O (1$-$0). 
% Similarly, in \citet{Zhou2022arXiv}, the hub-filament structure can be well identified by H$^{13}$CO+ (1$-$0).
The complete structure of W33 complex is far more extensive than the region marked by the blue box in Fig.\ref{vf} which is the mainly concerned area in previous works \citep{Immer2013-553,Immer2014-572,Kohno2018-70S, Murase2022-510,Tursun2022-658}. Fig.6 of \citet{Umemoto2017-69} shows the velocity range of $^{13}$CO (1$-$0) emission in W33 complex is (0, 60) km~s$^{-1}$. From the velocity channel maps in \citet{Dewangan2020-496}, excluding the low velocity components attributed to the background clouds, there are two velocity components [47.2, 60.2] km~s$^{-1}$ (called W33-red in this paper) and [29.6, 43.3] km~s$^{-1}$ (called W33-blue in this paper) which can represent two colliding clouds as suggested by \citet{Kohno2018-70S} and \citet{Dewangan2020-496}, also shown in Fig.\ref{vf}(a), Fig.\ref{bts}(a) and Fig.\ref{grid}(d). This paper only focuses on the [29.6, 43.3] km~s$^{-1}$ velocity component treated as a compression layer of cloud-cloud collision \citep{Kohno2018-70S}. From Fig.\ref{vf}, we can see a good hub-filament structure of W33-blue, which has also been noticed in \citet{Liu2021-646}.
% Almost the same velocity range has also been discussed carefully in \citet{Liu2021-646}, and a hub-filament system also been identified. 
In Fig.2 of \citet{Zhou2022-514}, the extinction map of \emph{Spitzer} 8 $\mu$m emission can well trace the hub-filament structure. Unlike \citet{Liu2021-646}, here we consider a more extended 8 $\mu$m extinction structure shown in Fig.\ref{vf}(b), where the center coordinate (l, b)=(12.9$^\circ$, -0.2$^\circ$), the width = $1.3^\circ$ and the height= $0.8^\circ$. The central coordinate of W33-main is (l, b)=(12.8$^\circ$, -0.2$^\circ$), marked by bold "+" in Fig.\ref{vf}(a). The spatial coverage of W33 complex estimated from 8 $\mu$m extinction is close to the investigated area in \citet{Dewangan2020-496}. Moreover, the low temperature and the high column density structures shown in Fig.\ref{t-n} can also keep in line with the morphology of hub-filament structure of W33-blue. 

Infrared dark clouds (IRDCs) and ATLASGAL clumps are selected from Table.1 of \citet{Peretto2016-590} and Table.5 of \citet{Urquhart2018-473}, respectively. We extract 88 clumps associated with W33 complex based on their distance close to 2.57 kpc firstly, then separate out 65 clumps according to
their systemic velocities within the range [29.6, 43.3] km~s$^{-1}$ and their location into the W33-blue region.
% firstly, then 65 clumps located in W33-blue are obtained by limiting $v\rm_{lsr}$ in the range [29.6, 43.3] km~s$^{-1}$. 
Most of ATLASGAL clumps in W33 complex are found in W33-blue, 
which could be a
result of the enhanced mass accumulation and subsequent fragmentation by the major compression layer in cloud-cloud collision.
% further confirms W33-blue is the main compression layer in cloud-cloud collision. 
There is no distance and velocity information of IRDCs, we extract the average velocity of each IRDC from the moment 1 map of $^{13}$CO (1-0) based on the central coordinate and radius listed in the Table.1 of \citet{Peretto2016-590}, and retain 113 IRDCs in the velocity range [29.6, 43.3] km~s$^{-1}$. All of these clumps and IRDCs in [29.6, 43.3] km~s$^{-1}$ are marked in Fig.\ref{vf}(b).

% Fig.\ref{channel} shows the integrated velocity channel maps of centroid velocity cube output by BTS fitting, we can see a continuous velocity distribution gradually transition from upper right to lower left following the increase of velocity, which is consistent with the moment-1 map of W33-blue in Fig.\ref{vf}(c). 

% In the following analyses, we mainly focus on the $^{13}$CO (1$-$0) emission, which can trace the whole hub-filament structure of W33-blue. More optically thin property of C$^{18}$O (1$-$0) and higher velocity resolution of $^{13}$CO (2$-$1) make these two molecular lines can provide good comparison with $^{13}$CO (1$-$0). 
% Tab.\ref{ingredient} lists the basic ingredients of this paper.

\subsection{Velocity structure in W33 complex}\label{single}
\begin{figure}
  \includegraphics[width=0.47\textwidth]{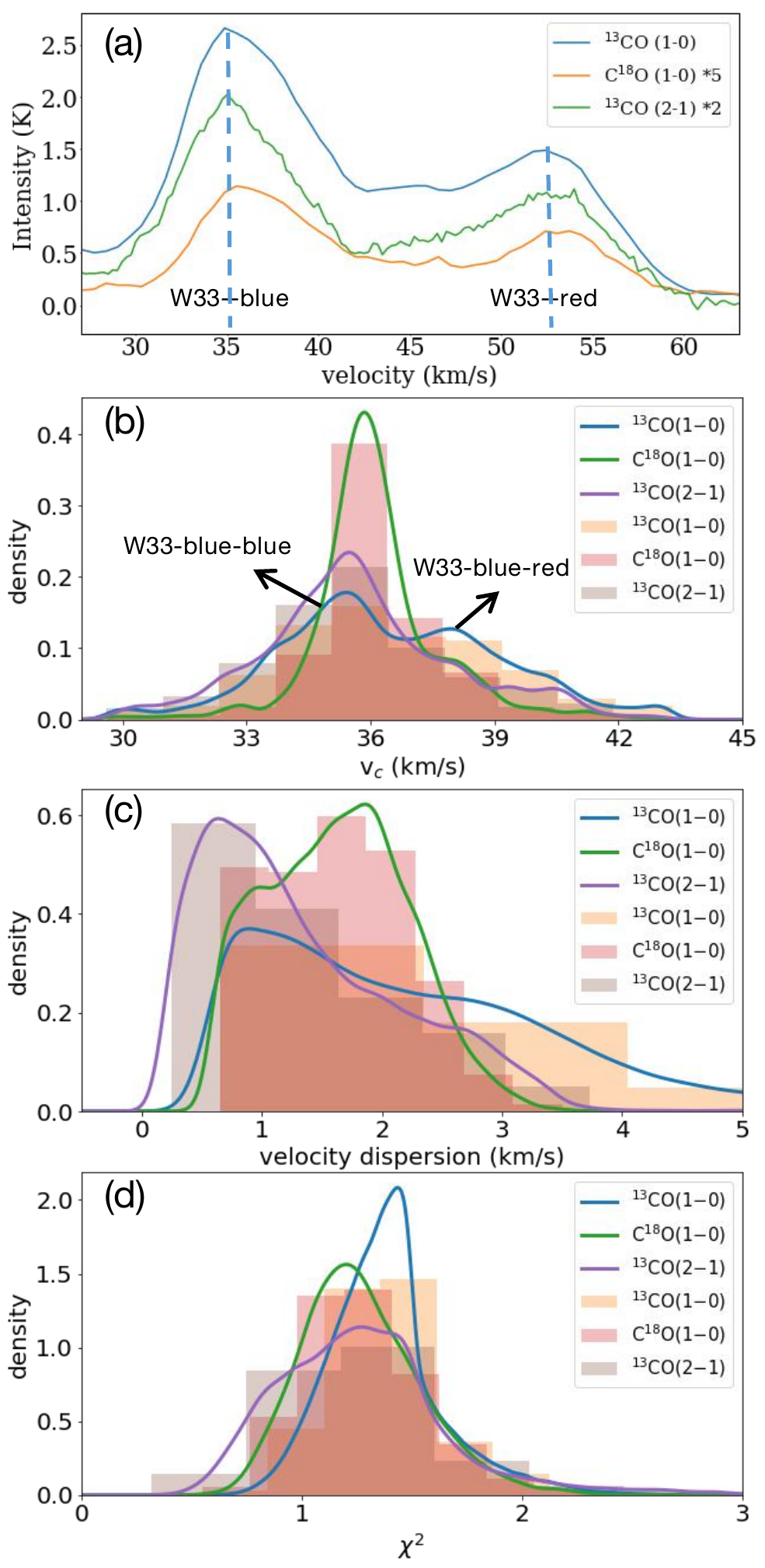}
\caption{(a) Average spectra of $^{13}$CO (1-0), C$^{18}$O (1-0) and $^{13}$CO (2-1) for the entire W33 complex;
(b), (c) and (d) Statistical distributions of centroid velocity, velocity dispersion and $\chi^2\rm_c$ value of $^{13}$CO (1-0), C$^{18}$O (1-0) and $^{13}$CO (2-1) line emission fitted by BTS algorithm for W33-blue.}
\label{bts}
\end{figure}

We use a fully automated multiple-component fitter Behind the Spectrum (BTS)\footnote{https://github.com/SeamusClarke/BTS} \citep{Clarke2018-479} to pixel-by-pixel fit the spectra of $^{13}$CO (1-0), C$^{18}$O (1-0) and $^{13}$CO (2-1) cubes in the velocity range [29.6, 43.3] km~s$^{-1}$. The integrated emission limit $5\sigma$ is used to determine if a spectrum is significant enough to warrant a good fitting performance, and we set the noise level as $1\sigma$. For each component, the BTS returns the peak intensity, centroid velocity (v$_{c}$), velocity dispersion ($\sigma_{c}$), and the reduced $\chi^2\rm_c$ value. $\chi^2\rm_c$ is used to determine the goodness of the fitting. All of these parameters are output in the position-position-velocity space as data cubes.

From Fig.\ref{bts}, the velocity distribution of $^{13}$CO (1$-$0) in W33-blue has two main peaks, indicating there are two velocity components shown in Fig.\ref{vf}(c), i.e. W33-blue-blue and W33-blue-red. In Fig.\ref{vf}, Fig.\ref{bts} and Fig.\ref{lte}, the emission of C$^{18}$O (1$-$0) and $^{13}$CO (2$-$1) mainly concentrates in the blue box marked in Fig.\ref{vf} 
that has the
strongest star formation activity in W33 complex, and corresponds to the W33-blue-blue velocity component.
% where has the strongest star formation activity in W33 complex, and this region is located in W33-blue-blue. 
The main peaks of C$^{18}$O (1$-$0) and $^{13}$CO (2$-$1) in Fig.\ref{bts}(b) also correspond well to the blueshifted peak of $^{13}$CO (1$-$0) emission. Thus we further speculate two velocity components traced by $^{13}$CO (1$-$0) are approximately independent. 
But together, they constitute a continuous velocity field, which can also be seen in the channel maps of \citet{Dewangan2020-496}. 
The continuous and monotonic velocity gradients in Fig.\ref{vg} can also reflect this fact, similar results also shown in the Fig.11 of \citet{Liu2021-646}. 
Fig.\ref{grid} shows the grid maps of three molecular lines used in this work, in the main emission regions, most of the spectral lines show single-peak profiles, 
especially for the filament structures. 
% The complex line profiles of $^{13}$CO (1$-$0) and $^{13}$CO (2$-$1) emission in several local regions may result from the relatively high optical depth shown in Fig.\ref{lte} or potential multiple velocity components.

Fig.\ref{3d} displays the centroid velocity distribution of BTS fitting in PPV space. We can see a single and continuous velocity field from W33-blue-red to W33-blue-blue, 
it’s consistent with the southeast-northwest gradient in the moment-1 map of Fig.\ref{vf}(c). 

\subsection{Velocity gradient}\label{gradient}
\begin{figure*}
  \includegraphics[width=0.8\textwidth]{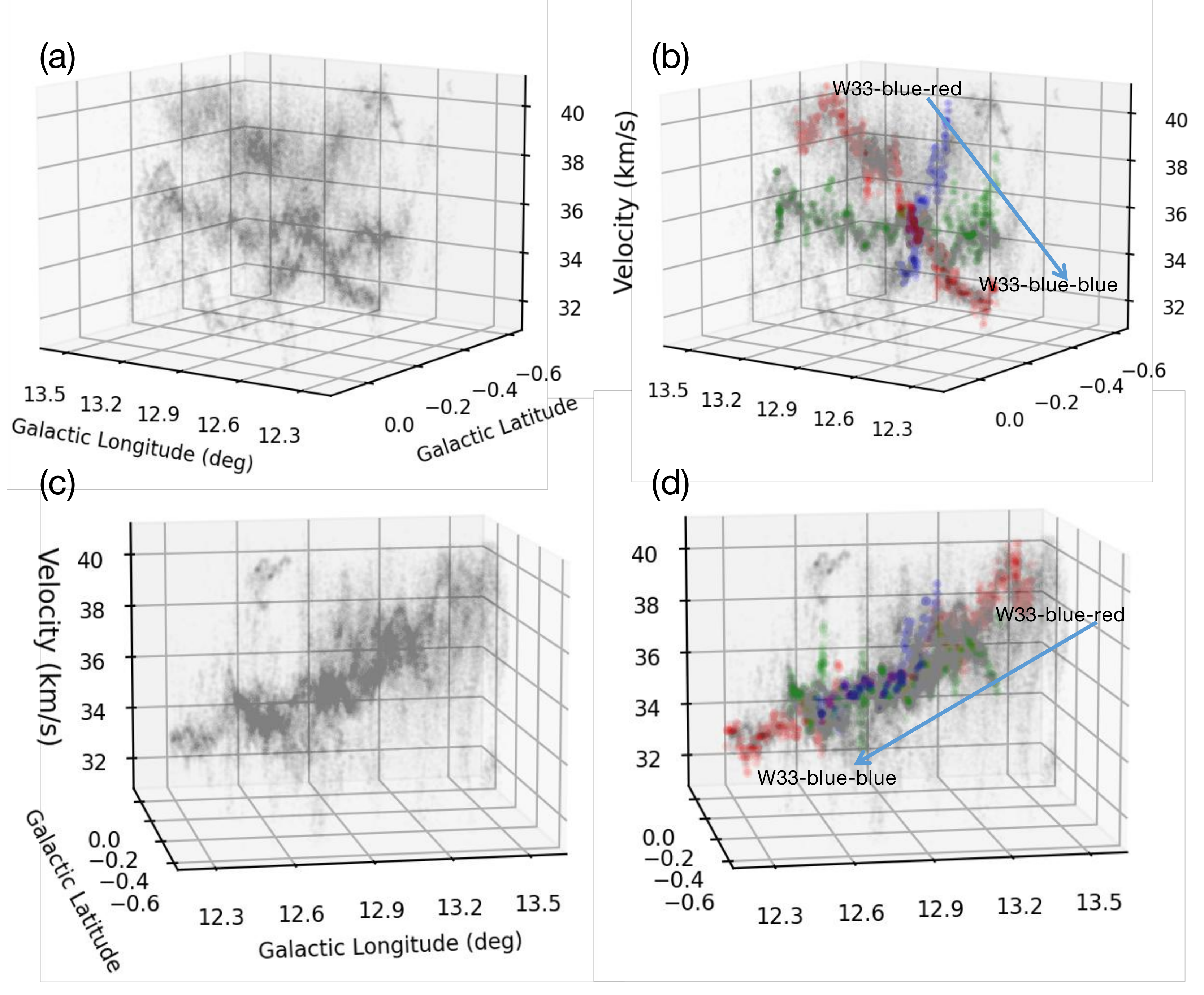}
\caption{The centroid velocity distribution in PPV space for the BTS fitting of $^{13}$CO (1$-$0) emission  viewed at different angles. Color dots show the velocity fields extracted from moment-1 map along the filament skeletons marked in Fig.\ref{vf}. Red dots represent f1\&f7, green dots show f3\&f4\&f5, blue dots are f6\&f2. Detailed distribution of the velocity and intensity along these filaments present in Fig.\ref{vg}.}
\label{3d}
\end{figure*}

In Fig.\ref{vf}(c), there is a clear velocity gradient across W33-blue, similar velocity mode is also found in G305 complex \citep{Hindson2013-435}, which may indicate that the cloud complex is not being viewed straight on but projected at an angle. At the interface between the ionization front and the molecular gas, the FUV photons from stars will excite polycyclic aromatic hydrocarbons (PAHs) on the surface of dense molecular clouds \citep{Tielens2008-46}. As discussed in \citet{Hindson2013-435}, we would expect the PAH and molecular emission to be superimposed upon each other if the orientation of W33-blue is at some inclination angle. If there is no inclination, the boundary between the PAH and molecular emission should be clear. Three previously identified O4–7(super)-giant stars around W33 Main in \citet{Messineo2015-805} are marked in Fig.\ref{vf}(d) and (f) by black "+". We can see the strong \emph{Spitzer} 8 $\mu$m emission likely excited by two massive stars just distributes along the boundary of $^{13}$CO (1$-$0) and (2$-$1) emission. Moreover, another star corresponds to the central cavity of $^{13}$CO (1$-$0) integrated intensity map, where also has strong \emph{Spitzer} 8 $\mu$m emission. These signatures mean that the projection effect of W33-blue may be not serious, or the projection may be not the only reason to cause large-scale velocity gradient in W33-blue.
\citet{Liu2021-646} has excluded the possibility that 
velocity gradients are from cloud rotation.
% From Fig.11 of \citet{Liu2021-646} and Fig.\ref{vg}, there are only monotonic velocity gradients, no signs of a Keplerian rotation signature means velocity gradients are not from cloud rotation.

The velocity gradient due to the longitudinal inflow is a basic feature of hub-filament system. \citet{Liu2021-646} has roughly estimated the velocity gradients along the similar directions of several filaments marked in Fig.\ref{vf}. Adopting the same method as in \citet{Zhou2022-514}, we use the FILFINDER algorithm \citep{Koch2015-452} to identify filaments from the moment 0 maps of $^{13}$CO (1$-$0). The filament skeletons shown in Fig.\ref{vf}(d) and (e) are highly consistent with the gas structures traced by $^{13}$CO (1$-$0) as seen by eye. Then we extract the velocity and intensity along the filament skeletons from the moment 0 and moment 1 maps of $^{13}$CO (1$-$0). In Fig.\ref{3d}, the velocity variations along the filaments extracted from moment-1 map (colorful dots) keep pace with the centroid velocity distribution of BTS fitting,  also indicating there is no the issue of multiple velocity components in W33-blue, thus the velocity gradients shown in  Fig.\ref{vg} estimated from moment-1 maps are reliable.
% Fig.\ref{vg} displays the distribution of velocity and intensity. 
We also check the velocity distribution along the filament skeletons using SEDIGISM $^{13}$CO (2$-$1) emission which has better velocity resolution than FUGIN $^{13}$CO (1$-$0) data. As shown in Fig.\ref{vg}, the velocity fields along the skeletons traced by $^{13}$CO (1$-$0) and $^{13}$CO (2$-$1) emission are very similar.
V-shape velocity structure around the peak of intensity or density indicates an accelerated material inflowing towards the central hub \citep{Gomez2014-791,
Kuznetsova2015-815, Hacar2017-602, Kuznetsova2018-473, Zhou2022-514}, however, there is mainly a single velocity gradient dominating the global velocity structure of W33-blue shown in Fig.\ref{vf}(c).
Although there are intense velocity fluctuations and many local velocity shears in Fig.\ref{3d} and Fig.\ref{vg}, the expected maximum velocity shear around the hub does not appear.
Moreover, the filaments f3 and f4 along the interface of W33-blue-blue and W33-blue-red do not manifest clear velocity gradients, indicating the velocity gradients of filaments f1\&f7 and f6\&f2 are mainly caused by different velocity components at large scale (i.e. W33-blue-blue and W33-blue-red), see Sec.\ref{origin} for more discussion.
% Thus filamentary structures in W33-blue may not represent the accretion flow towards the central hub. 
% thus we can exclude the influence of the low velocity resolution of FUGIN data. 

\subsection{LTE analysis}
To calculate the column densities of $^{13}$CO and C$^{18}$O, we assume conditions of local thermodynamic equilibrium (LTE) and a beam filling factor of 1. Following the procedures described in \citet{Garden1991-374,Sanhueza2012-756,
Mangum2015-127,Nishimura2015-216, Liu20, Areal2019-36, Liu2021-646, Menon2021-500}, for a rotational transition from upper level J + 1 to lower level J, we can derive the total column density by:
\begin{equation}
 N_{tot} =\frac{3k}{8\pi^3{\mu}^2B(J+1)}\frac{(T_{\rm ex}+hB/3k)\exp[hBJ(J+1)/kT_{\rm ex})]}{1-\exp(-h\nu/kT_{\rm ex})}
 \int{\tau{\rm dv}},
\label{density}
\end{equation}
% \begin{equation}
%   \tau = - ln(1 - T_{mb} [\frac{h\nu}{k}(\frac{1}{e^{h\nu/kT_{ex}}-1} - \frac{1}{e^{h\nu/kT_{bg}}-1})]^{-1})  
% \label{tau}
% \end{equation}
\begin{equation}
\tau ={\rm-ln}[1-\frac{T_{\rm mb}}{J(T_{\rm ex})-J(T_{\rm bg})}],
\end{equation}

\begin{equation}
\int{\tau{\rm dv}} = \frac{1}{J(T_{ex}) - J(T_{bg})} \frac{\tau}{1-e^{-\tau}}
\int{T_{\rm mb}{\rm dv}}
\label{tem}
\end{equation}

\begin{equation}
J(T) = \frac{h\nu/k}{e^{h\nu/kT}-1}
\label{j}
\end{equation}
% \frac{1}{J(T_{ex})-J(T_{bg})}\frac{\tau}{1-\exp(-\tau)} W, $E\rm_u$ is the energy of the upper level,
where $B=\nu/[2(J+1)]$ is the rotational constant of the molecule, $\rm \mu$ is the permanent dipole moment, and $\rm \mu= 0.112$ Debye and 0.1098 Debye for $^{13}$CO and C$^{18}$O.  $\rm T_{bg}=2.73$ is the background temperature, $\int{T_{\rm mb}{\rm dv}}$ represents the integrated intensity. In the above formulas, the correction for high optical depth was applied \citep{Frerking1982-262,Areal2019-36}, which is indeed required in the case of $^{13}$CO, but not for C$^{18}$O because it is mostly optically thin, as shown in Fig.\ref{lte}. 
% The specific 
% equations for each of the molecular spectral lines are listed in Appendix.\ref{lte-e}.
% Assuming the $^{12}$CO emission is optically thick and also in local thermodynamic equilibrium (LTE), 
% for each fit of $^{13}$CO and C$^{18}$O respectively via the observed $^{12}$CO peak brightness temperature $^{12}T_{\rm peak}$ 
Assuming optically-thick emission of $^{12}$CO (1-0), 
we can estimate the excitation temperature $T_{\rm ex}$ flowing the formula \citep{Garden1991-374,Pineda2008-679}
\begin{equation}
T_{ex} = \frac{5.53}{{\rm ln}[1 + 5.53 / (^{12}T_{\rm peak} + 0.818)]},
\label{tex}
\end{equation}
where $^{12}T_{\rm peak}$ is the observed $^{12}$CO (1-0) peak brightness temperature.

The excitation temperature distribution in Fig.\ref{lte}(g) is similar to the dust temperature distribution derived from \emph{Herschel} observation shown in Fig.\ref{t-n}. Due to the lack of $^{12}$CO (2-1) emission, we use the excitation temperature of $^{13}$CO (1-0) to derive the column density and optical depth of $^{13}$CO (2-1). The temperature range $\sim$10--30 K in Fig.\ref{lte}(g) is close to the LTE calculation of $^{13}$CO (2-1) in \citet{Menon2021-500}.

% \subsection{Projection and velocity gradient}
\subsection{Kinematics and dynamics of Dendrogram structures}
\subsubsection{Dendrogram structure identification}\label{deg}

Using {\it astrodendro} package \footnote{\url{https://dendrograms.readthedocs.io/en/stable/index.html}},
{\it Dendrogram} algorithm was conducted to extract the density structures of different scales from $^{13}$CO (1$-$0), C$^{18}$O (1$-$0) and $^{13}$CO (2$-$1) emission in the position-position-velocity (PPV) space. As described in \citet{Rosolowsky2008-679}, it decomposes the intensity data cube into hierarchical structures called leaves, branches and trunks. Leaves are defined as small-scale, bright structures at the top of the tree that do not decompose into further substructures. Branches are the larger scale structures in the tree, and they can break up into substructures. Trunks are the largest continuous structure at the bottom of hierarchical structures and, by definition, can also be single isolated leaves without any parent structure \citep{Mazumdar2021-656}. 
Three major parameters in this algorithm we used are {\it min\_value} for the minimum
value to be considered in the dataset, {\it min\_delta} for a leaf that can be considered as an independent entity, and {\it min\_npix} for the minimum area of a structure that is required to have an area equal to the area of the beam in our calculation.
For the strong $^{13}$CO (1$-$0) emission, we take {\it min\_value}~$=5 \sigma$ and {\it min\_delta}~$=2 \sigma$. For the relatively weak C$^{18}$O (1$-$0) and $^{13}$CO (2$-$1) emission, these two values are 5 $\sigma$ and 1 $\sigma$, respectively.
% Finally, we obtain 2979 structures in $^{13}$CO(J =1$-$0) cube and 235 structures in C$^{18}$O(J =1$-$0) cube, because the emission area of C$^{18}$O(J =1$-$0) is far less than $^{13}$CO(J =1$-$0). In Fig.\ref{}, $^{13}$CO(J =1$-$0) is optically thick in some regions, which may influence the estimation of velocity dispersion, thus the comparison with C$^{18}$O(J =1$-$0) is necessary. From Fig.\ref{tree}, there are at least four segments due to the highly developed branch structures shown in Fig.\ref{vf}, or they are identified in different velocity channels. 

{\it Dendrogram}  can create a catalog of some basic properties of all the structures, such as their positions, mean and RMS velocities (v$_{\rm rms}$), total fluxes and the projections of the projected major and minor axes (called $\sigma_{\text{maj}}$ and $\sigma_{\text{min}}$) on the position-position plane. We discard the isolated structures located near the edges before calculating the structure properties. Following \citet{Rosolowsky2006-118}, the physical radius of the structure was calculated by R$_{\text{pc}}$ =
R$_{eff}$*d = $\eta*\sqrt{\sigma_{\text{maj}}*\sigma_{\text{min}}}$*d, here $\eta=1.91$, $d=2.4~{\rm kpc}$ for the distance to W33 complex \citep{Immer2013-553}. Mass of each structure was calculated using the extracted column density according to the central coordinate and effective radius
\begin{equation} %\label{eq:1}
M = \mu_{\rm H_{2}} m_{\rm H} X_{\rm ^{13}CO}^{-1} N_{\rm ^{13}CO} \, A,
\label{mass}
\end{equation}
where $\mu_{\rm H_{2}} = 2.8$ is the molecular weight per hydrogen molecule, 
$m_{\rm H}$ is the hydrogen atom mass. The abundance ratio X$_{\rm ^{13}CO}$ of H$_2$ compared with $^{13}$CO is $\sim 7.1 \times 10^5$ \citep{Frerking1982-262}, and A is the area of each structure, here we take A $\approx \pi$R$_{\text{pc}}^{2}$ . Moreover, the dynamical state of clump can be captured by the virial parameter \citep{Bertoldi1992}, defined as 
\begin{equation}
 \alpha \equiv 5\sigma_{tot}^2R/(GM) = 2aE_K/|E_G|, 
\label{alpha}
\end{equation}
\begin{equation}
  \sigma_{tot} = \sqrt{\sigma_{\rm nt}^2+\sigma_{\rm t}^2}
 = \sqrt{\sigma^2-\frac{kT}{\mu_{\rm A} m_H}+\frac{kT}{\mu_p m_H}},
\label{viril-d}
\end{equation}
where $\sigma_{tot}$ and $R$ are the total 1D  velocity dispersion and radius of the structure. $\mu_p=2.37$ is the mean molecular weight per free particle, $\mu_{\rm A}$ is the molecular weight equal to 29 for $^{13}$CO and 30 for C$^{18}$O. The dimensionless parameter $a$ accounts for modifications that apply in the case of non-homogeneous and non-spherical density distributions. For a self-gravitating, unmagnetized clump without rotation, a virial parameter above a critical value $\alpha=2a$ indicates that the clump is unbound and may expand, while one below $2a$ suggests that the clump is bound and may collapse. For a spherical clump with a radial density profile that is a power law $\rho\propto r^{-k_\rho}$, then for $k_\rho=0,1,1.5,2$, $a=1,10/9,5/4,5/3$ \citep{Cheng2020-894}. We adopt a fiducial value of $a=1$. 

In Fig.\ref{co}, mean velocity distribution peaks are around the systematic velocity, indicating that most of {\it Dendrogram} structures contain representative kinematic information of the entire region. The spread of mean velocity agrees with the non-uniform velocity field of W33-blue, see Fig.\ref{vf}(c). Greater v$_{\rm rms}$ values of branches could be related to turbulence, gravity-driven motions, or large-scale ordered motions (such as outflow and cloud-cloud collision). 
% which is also consistent with the Larson-relation. 

% \subsection{Velocity dispersion and column density}\label{velocity-column}
\subsubsection{Several key physical quantities}\label{key}
Before making quantitative calculations, we need to clarify several key physical quantities:
1. $\sigma_{\rm M2}$, velocity dispersion extracted from moment-2 map according to the central coordinate and effective radius of each {\it Dendrogram} structure; 2. N$_{\rm LTE}$, extracted from the column density map derived from LTE analysis according to the central coordinate and effective radius of each {\it Dendrogram} structure; 3. N$_{\rm PPMAP}$, extracted from the column density map derived from Hi-GAL data processed by PPMAP procedure according to the central coordinate and effective radius of each {\it Dendrogram} structure; 4. v$_{\rm rms}$, velocity dispersion output by {\it Dendrogram} algorithm; 5. $\sigma_{\rm BTS}$, velocity dispersion extracted from the velocity dispersion cube fitted by BTS algorithm. Each {\it Dendrogram} structure identified in PPV space has the central coordinate (v$_{\rm cen}$, y$_{\rm cen}$, x$_{\rm cen}$) and radius (v$_{\rm rms}$, R$_{\rm eff}$, R$_{\rm eff}$), which can be used to obtain the 3D slice for each structure in the velocity dispersion cube. We can calculate the average velocity dispersion of each structure in the slice, which can effectively separate the overlapping of multiple velocity components.
% Moment-2 maps of molecular lines () and v$_{rms}$ are calculated by the same formula \citep{Rosolowsky2006-118}
% \begin{eqnarray}
% \sigma_v & = &\sqrt{\sum_{i} T_i \left[ v_i -
% \bar{v} \right]^2/\sum_{i} T_i} \\
% \bar{v}& = & \sum_{i} T_i v_i/\sum_{i}
% T_i \mbox{ .}
% \label{m2}
% \end{eqnarray}
However, BTS is designed to only work for optically thin spectra as it assumes a Gaussian shape to the profile,  and thus it ought not be used on spectra dominated by non-Gaussian components, e.g. optically thick spectra or highly skewed spectra. In Fig.\ref{lte}, the optical depth of $^{13}$CO(1$-$0) and $^{13}$CO(2$-$1) are indeed larger than 1 in several local regions. The distribution of $\chi^2$ in Fig.\ref{bts} is similar to the Fig.6 of \citet{Clarke2018-479}, which also reveals the fitting of C$^{18}$O(1$-$0) is better than $^{13}$CO(1$-$0) and $^{13}$CO(2$-$1).

In Fig.\ref{co}, the values of $\sigma_{\rm BTS}$ are significantly greater than
$\sigma_{\rm M2}$ for all molecular lines. The discrepancy may be attributed to the difference of calculation methods. Moreover, there is no dependency between $\sigma_{\rm M2}$, $\sigma_{\rm BTS}$ and the scales. The major reason is that the velocity dispersion $\sigma_{\rm M2}$ and $\sigma_{\rm BTS}$ are calculated on the pixel scale, and thus the larger scale velocity structure of each {\it Dendrogram} structure will be removed from the velocity dispersion estimate.
Exactly, BTS algorithm is pixel-by-pixel fitting the spectra,  while Moment-2 map is calculated for each data point directly\footnote{\url{https://spectral-cube.readthedocs.io/en/latest/moments.html\#moment-maps}}, see Sec.\ref{app-A} for more detailed discussion about the difference of $\sigma_{\rm M2}$ and  $\sigma_{\rm BTS}$.
Although v$_{\rm rms}$ is also the intensity-weighted second moment of velocity, it is calculated for each complete {\it Dendrogram} structure (not for each pixel), thus only v$_{\rm rms}$ can be used to discuss the Larson-relation and calculate the virial parameters of each {\it Dendrogram} structure. 
However, kinematic information output by {\it Dendrogram} is provided in the form of intensity-weighted average and rough quantities relating to each structure \citep{Henshaw2019-485}. If we are interested in how those kinematic quantities vary with position within a given structure on the pixel scale, $\sigma_{\rm M2}$ and $\sigma_{\rm BTS}$ can roughly meet this requirement. For the small scale structures (leaves), considering the CO depletion and the uncertainty of abundance, it is necessary to compare the column density derived from LTE (N$_{\rm LTE}$) with that from Herschel observations (N$_{\rm PPMAP}$).
% comparing the column density derived from LTE (N$_{\rm LTE}$) and the column density derived from Herschel data (N$_{\rm PPMAP}$) is also necessary. 

% The distribution of N$_{PPMAP}$ is more concentrated. In Fig.\ref{13co} and 
% Fig.\ref{co}, overall, the column density N$_{LTE}$ derived from C$^{18}$O (1-0), $^{13}$CO (2-1) and $^{13}$CO (1-0) is larger, smaller and similar than N$_{PPMAP}$, respectively. 

Deriving the column density from LTE, SED or PPMAP methods is carried on the PP maps, the integrated intensity inherently contains the contribution of potential multiple velocity components, thus $\sigma_{\rm M2}$ derived from moment-2 map has the same prerequisite with the column density (i.e. homologous) compared with $\sigma_{\rm BTS}$ and v$_{\rm rms}$ derived from PPV space. Actually, in the LTE analysis of this paper, the integrated intensity is from the moment-0 map. It may be the reason why the column density has a stronger correlation with $\sigma_{\rm M2}$ shown in Fig.\ref{sig-n} and Fig.\ref{sig-nn}, which should not be simply taken as a coincidence. $\sigma_{\rm M2}$ can match with N$_{\rm PPMAP}$ and N$_{\rm LTE}$ better than $\sigma_{\rm BTS}$ and v$_{\rm rms}$, although all of $\sigma_{\rm M2}$, N$_{\rm PPMAP}$ and N$_{\rm LTE}$ can't exclude the issues of potential multiple velocity components. In short, we should treat three kinds of velocity dispersion $\sigma_{\rm M2}$, $\sigma_{\rm BTS}$ and v$_{\rm rms}$ carefully, they may reflect different physics, rather than simply the pros or cons or alternative relationships, also refer to  \citealt{Liu22}. In principle, {\it the physical quantities estimated from PPV space and PP plane shouldn't be mixed to calculate other physical quantities}. But in the case of W33-blue, the effects of potential multiple velocity components can be ignored, as discussed in Sec.\ref{single}. 

\subsubsection{Virial state}
Velocity dispersion v$_{\rm rms}$ and two kinds of column density N$_{\rm PPMAP}$ and N$_{\rm LTE}$ are used to calculate the virial parameter. As shown in Fig.\ref{co}, most of structures are gravitationally bounded, indicating the dominance of gravity in W33-blue.
The values of virial parameter increase with the decreasing of scales, indicating there may be hierarchical “self-gravitating cocoons” from large scale to small scale. This scenario has been discussed in \citet{Rosolowsky2008-679, Goodman2009-692}, they emphasized the importance of gravity on all possible scales, but self-gravitating structures are more prevalent on larger scales than on smaller ones. However, it doesn't mean the structures at small scales are more gravitationally unbound, because external pressure from the gravitational collapse of large scale structures may play an important role to bound the small-scale structures. The additional binding pressure from ongoing gravitational collapse is different from the scenario that the external pressure provided by the ambient molecular cloud materials might help to confine dense structures \citep{Elmegreen1989-338,Field2008-385,Kirk2017-846,Li2020-896}. Although the latter may also contribute to the binding pressure. Thus the virial analysis considering only the kinetic and gravitational energies may be not applicable for small-scale structures.

\subsubsection{Dependencies of $\sigma_{\rm M2}$ and $N_{\rm LTE}$}

\begin{figure}
  \includegraphics[width=0.5\textwidth]{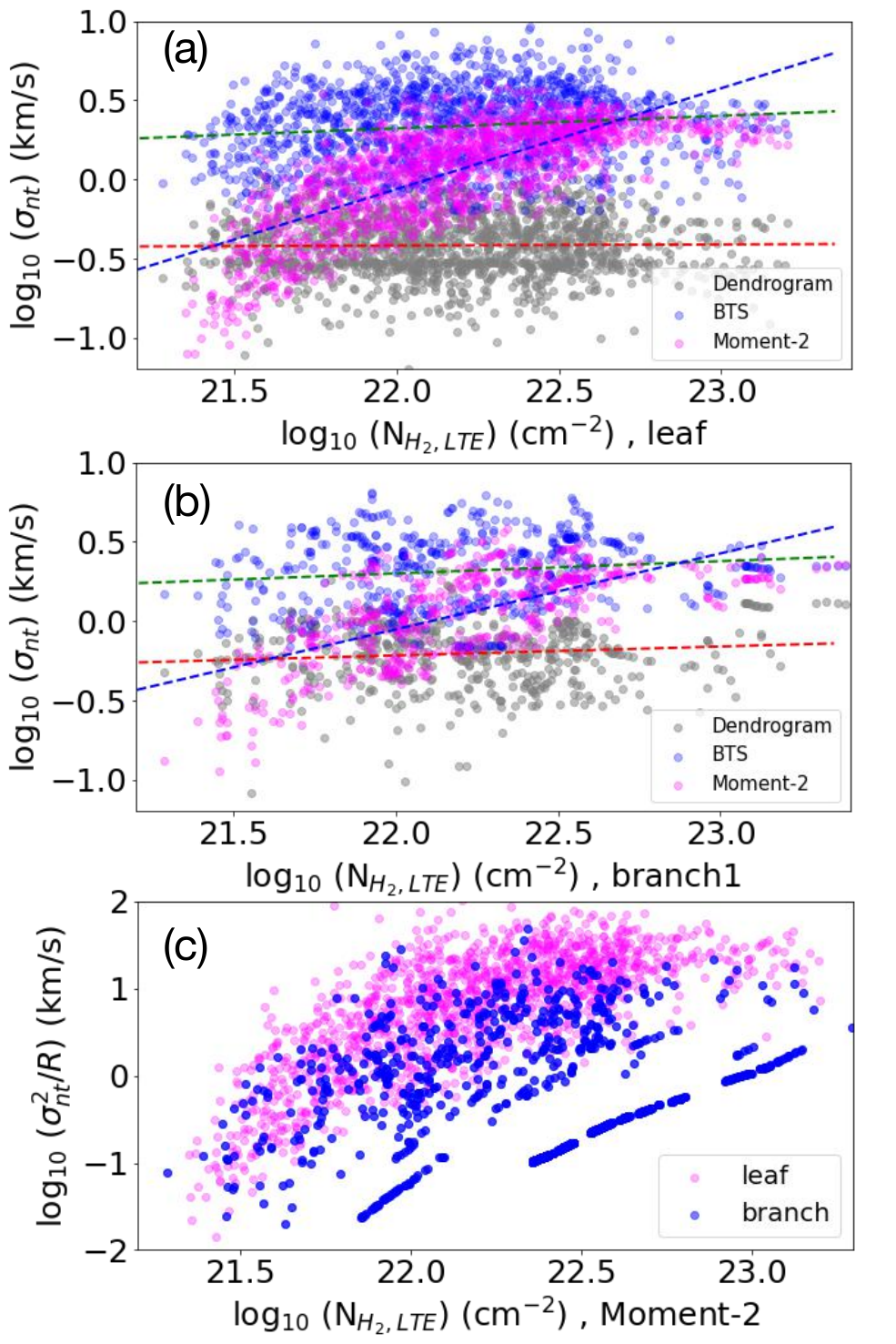}
\caption{(a) $\sigma_{nt}$--$N_{\rm H_2, LTE}$ relation of leaf structures. Gray, blue and magenta points represent v$_{\rm rms}$, $\sigma_{\rm BTS}$ and $\sigma_{\rm M2}$, respectively; (b) $\sigma_{nt}$--$N_{\rm H_2, LTE}$ relation of branch structures; (c) $\sigma_{nt}^2/R$--$N_{\rm H_2, LTE}$ relation, i.e. $a_{\mathrm{G}}$-- $a_{k}$ relation, here $\sigma_{nt}$ is estimated from $\sigma_{\rm M2}$. Blue and magenta points represent the branch and leaf structures.}
\label{sig-n}
\end{figure}

Although $\sigma_{M2}$ and N$_{LTE}$ can't exclude the potential multiple velocity components, they are the only two homologous physical quantities and show strong dependency in Fig.\ref{sig-n} and Fig.\ref{sig-nn}. The assumption of gravity driving the chaotic motions in multiple local centres of collapse satisfy 
\begin{equation}
  \sigma_{nt} \propto \sqrt{\Sigma R},
  \label{larson_g}
\end{equation}
where $\sigma_{nt}$, $\Sigma$ and $R$ represent the non-thermal velocity dispersion, surface density and the size of the structure. It implies that massive, compact structures should develop larger velocity dispersion for larger column density. The local gravitational contraction occurs throughout the whole cloud, which is itself collapsing, thus this relation to be valid not only for local structures, but also for molecular clouds in general \citep{Ballesteros2011-411,Ballesteros2018-479,Traficante2018-473,Vazquez2019-490}. In this scenario, the hierarchical and chaotic gravitational collapse can generate the supersonic linewidths in molecular clouds, rather than hydrodynamical turbulent motions. Equivalently, if the observed regions at increasing surface densities $\Sigma$ or gravitational acceleration $a_{\mathrm{G}}$ (due to $a_{\mathrm{G}}=\pi G\Sigma/5$) , and we can observe an increase of kinetic acceleration $a_{k}=\sigma_{nt}^{2}/\mathrm{R}$ from the non-thermal motions, it also suggests that on average the majority of the non-thermal motions originate from gravitationally driven chaotic collapse \citep{Traficante2018-473}. The strong positive correlations of $\sigma_{nt}$--$N_{\rm H_2}$ and $a_{\mathrm{G}}$-- $a_{k}$ are indeed shown in Fig.\ref{sig-n} and Fig.\ref{sig-nn}.

\subsection{Gravitational stability of filaments}\label{stability}
\begin{table}
	\centering
	\caption{$\sigma_{\rm V}$, velocity dispersion of the filament or sub-region marked in Fig.\ref{vf}(d); M, mass of the filament; L, length of the filament; f, ratio of the line mass to a critical line mass.}
	\label{width-pdf}
	\begin{tabular}{ccccccc} % five columns, alignment for each
	    \hline
	region	&	$\sigma_{\rm V}$	&	M& L& f\\
    &(km/s) &($\mathrm{M}_\odot)$ &(pc) & \\
	\hline
    2	&	2.72 $\pm$ 0.15 	&	4.33 $\times$ 10$^{4}$& 11.72 &1.19 \\
    3	&	3.75 $\pm$ 0.09 	&	 9.35 $\times$ 10$^{4}$& 14.65& 1.03 	\\
    4	&	4.92 $\pm$ 0.05 	&	 2.49 $\times$ 10$^{4}$&  -- &-- 	\\
    5	&	1.45 $\pm$ 0.06 	&	 2.82 $\times$ 10$^{4}$& 9.21& 3.26\\
    6	&	1.78 $\pm$ 0.08 	&	2.82 $\times$ 10$^{4}$& 16.74& 1.23\\
	\end{tabular}
\end{table}

Mass $M$ and length $L$ define a filament's line mass $m$, $m = M/L$. The critical line mass of a filament m$_\mathrm{crit}$(T)=$2 c_\mathrm{s}^2/G$, which is derived from a hydrostatic, isothermal cylinder model \citep{Stodolkiewicz1963-13,Ostriker1964-140}. To include the contribution from non-thermal motions to the filament's stability, the term $c_\textrm{s}$ can be replaced by the total velocity dispersion to obtain the virial line mass \citep{Wang2014-439,Hacar2022-arXiv}
\begin{equation}
    m_\mathrm{vir}(\sigma_\mathrm{tot})=\frac{2 \sigma_\mathrm{tot}^2}{G}\sim 465\left(\frac{\sigma_\mathrm{tot}}{1\,{\rm km \, s}^{-1}}\right)^2 \, \mathrm{M}_\odot \, \mathrm{pc}^{-1} \, .
\end{equation}
The ratio of the line mass to a critical line mass can be used to estimate the gravitational state of filaments \citep{Ostriker1964-140,Fischera2012-547}, $f\equiv m/m_\mathrm{crit}$. 
Filaments with $f>1$ (super-critical) become radially unstable and must collapse under their own gravity, while only those with $f<1$ (sub-critical) can remain in hydrostatic equilibrium.

Table.\ref{width-pdf} lists the velocity dispersion of each filament in W33-blue by Gaussian fitting to each sub-region marked in Fig.\ref{vf}(d). 
We further calculate the total velocity dispersion according to equation.\ref{viril-d}. The mass of each filament is estimated by 
\begin{equation}
M = \mu_{\rm H_2} {\rm m_H} \sum N{\rm(H_2)}(R\rm_{pixel})^2,
\end{equation}
where $ R\rm_{pixel}$ is the size of a pixel, $\rm \mu_{H_2}=2.8$ is the mean molecular weight per hydrogen molecule, and $\rm m_H$ is the mass of a hydrogen atom. Here, correctly measuring the lengths of filaments is difficult due to no clear boundary between these filaments and the hub region, also the irregular morphology of filaments. That's why we gave up sub-region 4. 
The values of $f$ also rely sensitively on the estimation of the velocity dispersion. Containing these significant uncertainties,
the final values of $f$ listed in Table.\ref{width-pdf} are close to the critical value ($\sim$ 1), which should be treated carefully. However, $f \sim 1$ is not absurd in the case of W33-blue, which indicates W33-blue is still very turbulent due to the ongoing cloud-cloud collision and also in the transition state from the compression layer to a hub-filament system, see Sec.\ref{hf-f} for more discussion.

\section{Discussion}

\subsection{Origin of two velocity components}\label{origin}

In this paper, we only consider the gas within velocity range [29.6, 43.3] km s$^{-1}$ in W33 complex. As shown in Fig.\ref{vf}(c), there are still two velocity components in this velocity interval. In Fig.\ref{vf}(a), two clouds with different velocities has been attributed to cloud-cloud collision (CCC) scenario in \citet{Kohno2018-70S} and \citet{Dewangan2020-496}. Most of ATLASGAL clumps correspond to the high density part of Herschel H$_{2}$ column density well, both of them reveal bent structure marked by black dashed curve in Fig.\ref{vf}(a), and W33-main is located just at the bottom. Evidence has accumulated implying that the presence of compressed layer caused by two colliding clouds and then young stellar clusters and massive stars form at the shock-compressed interface layer \citep{Habe1992-44,Anathpindika2010-405,Inoue2013-774L,Takahira2014-792,Haworth2015-450,Torii2017-835,Bisbas2017-850,Kohno2018-70S}. 
Low temperature structure mainly settles down in W33-blue-red, the impact crater from CCC process is mainly located in W33-blue-blue and thus results in higher temperature. The following active star formation also contributes to the increase of temperature in W33-blue-blue. As shown in Fig.\ref{vf}, the boundary of two different velocity components is also the boundary of low and high temperature structures, and the main high density part of Herschel H$_{2}$ column density and most of ATLASGAL clumps are located along the boundary, thus the boundary could be the interface layer of collision.  
Thus we speculate the CCC process mainly happened in W33-blue-blue may be an important factor to cause the velocity difference (i.e. velocity gradient) in W33-blue. In the direct collision region, we can see more broken structures, 
% for example, a piece of filamentary cloud “f5” with relatively low density is flying away from the collision region shown in Fig.\ref{vf}(d), 
indicating the CCC process in W33 complex is still ongoing. Thus we can expect more intense star formation in the future with further coalescence of two colliding clouds. Moreover,
the following active star formation in W33-blue-blue makes its feedback effect stronger than W33-blue-red, which may be another significant factor to cause velocity difference in W33-blue. 
% Combined with the results in Sec.\ref{gradient}, 
% that there may be no detectable accretion flow along the filaments, % we conclude that velocity gradient in W33-blue is mainly from the projection, collision and feedback.

% \begin{figure*}
%   \includegraphics[width=0.65\textwidth]{w33/ccc.pdf}
% \caption{}
% \label{ccc}
% \end{figure*}
% \begin{figure*}
%   \includegraphics[width=0.65\textwidth]{w33/hf-t.pdf}
% \caption{}
% \label{hf-t}
% \end{figure*}

\subsection{Hub-filament structures triggered by cloud-cloud collision}\label{hf-f}

Bent structure shown in Fig.\ref{vf}(a) roughly corresponds to filaments f3 and f7 may indicate some filaments in W33-blue-blue directly formed in the intense collision process. However, in the compression scenario of \citet{Myers2009-700}, the initial clump which can generate the hub is an isothermal spheroid whose peak density is much greater than that of its surrounding uniform medium. As shown in Fig.\ref{vf}, filament f1\&f7 can span the entire structure of W33-blue, the column densities vary only a few times along the filament in Fig.\ref{vg}(c), which means the density distribution in the central hub and the filaments are comparable, this may be an evidence to demonstrate that W33-blue is still in the transition from the compressed layer to a hub-filament system.
% Considering the aforementioned dominance of gravity in W33-blue, 
Analyses in previous sections tend to support that W33-blue is dominated by gravity,
another reasonable choice is that the gravity quickly took over the formation of filaments after generating the compressed layer by collision, as described in some simulations and models \citep{Burkert2004-616, Gomez2014-791, Balfour2015-453}, see the description in Sec.\ref{introduction}.
% These filaments can be regarded as "splash" structures. 
Moreover, W33-blue-red is less influenced by the collision and feedback activity, but still has clear filamentary structures. These filaments may be generated more slowly by gravitational collapse, compared with quickly compressed.
Anyway, at present, the filamentary compressed layer W33-blue is in a state of gravitational instability, it may collapse into more complex filamentary structure in the future.

It is likely that formation of hub-filament structures in W33 complex is triggered by cloud-cloud collision. This mode may be very common in star formation regions, the collision generates dense compressed layer, then the dense layer will fragment or collapse into hub-filament structure. The subsequently formed hub-filament structure plays an important role in the transition from the compressed layer to a massive protocluster, because the hub-filament structure can convey materials by filaments to maintain the growth of protostars inside the central hub, the hub will evolve into protocluster finally.
The large systematic velocity difference of W33-blue ($\sim$ 35 km/s) and W33-red ($\sim$ 53 km/s) may indicate an ongoing cloud-cloud collision, if we assume a complete inelastic collision scenario, further collision and fusion of two clouds will trigger more intense star formation activity in the future. At present, W33-blue is still a very young star cluster, thus has extensive \emph{Spitzer} 8 $\mu$m extinction structure. It also suggests that
W33-blue may be still in the transition from the compressed layer to a hub-filament system. 
% the current velocity gradients from inflow are too small to be distinguished from other kinds of larger velocity gradients (such as the projection, collision and feedback), as discussed in Sec.\ref{gradient}.
% Now we can only determine that W33-blue is in a global collapse state dominated by gravity, from Fig.\ref{vf} and Fig.\ref{lte}, the central hub of W33-blue occupy most of the mass, 
In some theoretical models, the free-fall process is extremely nonlinear, proceeding very slowly at the beginning, and accelerating enormously towards the end \citep{Heitsch2008-689,Zamora2012-751,Zamora2014-793,Vazquez2019-490}, thus hierarchical and chaotic gravitational collapse comes to dominate at late stages. In Sec.\ref{stability}, the values of $f$ close to the critical value may imply a slow gravitational collapse at the beginning. 
As \citet{ Girichidis2014-781} note that, after 99\% of the collapse time, the radius has only decreased by one order of magnitude. In the SPH simulation of \citet{Gomez2014-791}, the larger-scale hub-filament system indeed forms towards the end of the simulation. In \citet{Peretto2022-257}, hub-hosting clumps are more massive and more evolved than non-hubs, the rapid global collapse of clumps is responsible for (re)organising filament networks into hubs. It implies that hub-filament systems may be the products when star formation regions evolve to a relatively late stage. Thus the dynamical features of W33-blue hub-filament system may be more pronounced as W33 complex evolves. 

In GHC model, the multi-scale gravitational collapse means local collapse centers will develop throughout the cloud while the cloud itself is also contracting, giving rise to chaotic density and velocity fields. The local collapse occurs due to the initial turbulence, in combination with strong radiative losses, inducing non-linear density fluctuations that have shorter free-fall times than the parent cloud \citep{Heitsch2008-689, Ballesteros2011-411, Vazquez2019-490}. Although there is no overall V-shape velocity structures around the main intensity peak along the filaments in W33-blue, we indeed find some local V-shape velocity structures around local intensity peaks in Fig.\ref{vg}, which may indicate the existence of local collapse centers. Infrared dark cloud SDC13 marked in Fig.\ref{vf}(b) can serve as an excellent representative of the local collapse center, which is also a typical case of relatively small scale hub-filament system \citep{Peretto2014-561}. As shown in Fig.\ref{vf}(b), IRDCs and ATLASGAL clumps mainly distribute in the extended 8 $\mu$m extinction structure. W33-blue as a huge hub-filament system can also be characterized by the 8 $\mu$m extinction. If IRDCs and ATLASGAL clumps in W33-blue can be treated as small scale hub-filament systems, there is a vivid hierarchical hub-filament structures in W33-blue as raised in \citet{Zhou2022-514}. \citet{Zhou2022-514} also found when the scale is less than $\sim$ 1~pc, gravity gradually dominates gas infall, thus the relatively small scale hub-filament structures (such as SDC13) can develop more easily than the large scale one (W33-blue). It may also mean a self-organization process from small scale hub-filament structures to the large scale one. Macroscopically ordered dynamical and kinematical structures at large scales are formed by the synergy between the small scale constituent units, this idea is worth doing for further verification.
% Moreover, the physical mechanism for forming hub-filament structure is very useful to distinguish between the different theoretical models, such as inertial inflow and GHC. 

In \citet{Williams2018-613}, large radial velocity gradients across two filaments of SDC13 cannot be due to gravity nor rotation, they propose a scenario for the evolution of the SDC13 hub-filament system in which filaments first form as post-shock structures in a supersonic turbulent flow. Then gravity takes over and starts shaping the evolution of the hub, both fragmenting filaments and pulling the gas towards the centre of the gravitational well. 
% By doing so, gravitational energy is converted into kinetic energy in both local (cores) and global (hub centre) potential well minima, generating more massive cores at the hub centre as a result of larger acceleration gradients. 
\citet{Wang2022-931} found that SDC13 is located at the convergent point of three giant molecular clouds (GMCs), combining large- and small-scale analyses, they propose a similar scenario with \citet{Williams2018-613}, SDC13 is initially formed from a collision of clouds moving along the large-scale magnetic field. In the post-shock regions, after the initial turbulent energy has dissipated, gravity takes over and starts to drive the gas accretion along the filaments toward the hub center. Combined with our study, 
W33-blue is the entire compression layer of cloud–cloud collision, SDC13 is only a little piece on the entire compression layer. Similar physical processes described in \citet{Williams2018-613} and \citet{Wang2022-931} can also occur throughout W33-blue, indicating the similar physics in hub-filament systems at different scales, i.e. hierarchical/multi-scale hub-filament structures.

% Turbulent compression and gravitational collapse, both of them have an important role in shaping the hub-filament structure. W33-blue is still in the early stage, with the end of the entire cloud-cloud collision process, where the two clouds completely merge, the whole structure will be dominated by gravity.
% Thus small-scale structures survive in strongly turbulent cloud collision environments.
% Thus although cloud-cloud collision causes intense turbulence in W33-blue, but the overall hub-filament structure and even the inner structures are still in a gravitationally bounded state, even the gravitational collapse. 
% Large scale structures are gravitationally bounded, thus the strong shock from cloud-cloud collision won't disperse W33-blue. The collision and turbulent compression can enhance the density, gravity plays a key role in shaping the morphology. In the models of Myers+2009 and Padoan+2020,

% gravity takes over earlier at small scale relative to large scale. global collapse with local gravitational center. shear, turbulence, gravity mix together, all of them participate in the star formation.
% Scenario: cloud-cloud collision, edge-on collapse, turbulence is not the primary one.

\section{Summary}
Using the survey data FUGIN $^{13}$CO (1$-$0), C$^{18}$O (1$-$0) and SEDIGISM $^{13}$CO (2-1), we investigated the gas kinematics and dynamics in W33 complex, and discussed the formation of hub-filament structure in W33-blue in detail, the main results of this work are as follows:

1. We estimate a more complete dynamical range of W33 complex from 8 $\mu$m extinction, The central coordinate (l, b)=(12.9$^\circ$, -0.2$^\circ$), width = $1.3^\circ$ and height= $0.8^\circ$. Combined with previous works, we divide W33 complex into W33-blue and W33-red which represent two colliding clouds in the cloud-cloud collision scenario. 74\% ATLASGAL clumps in W33 complex are located in W33-blue further confirms W33-blue exists as a compression layer. We fit the spectra of CO molecular line cubes pixel-by-pixel using multiple-component fitter BTS, the velocity distribution of $^{13}$CO (1$-$0) in W33-blue reveals a continuous, approximately single velocity field, together with the moment 1 map of $^{13}$CO (1$-$0), we further divide W33-blue into W33-blue-blue and W33-blue-red. The velocity components of C$^{18}$O (1$-$0) and $^{13}$CO (2$-$1) emission mainly concentrate in W33-blue-blue where has the strongest star formation activity in W33 complex and higher temperature than W33-blue-red.  

2. We derive the column density N$_{\rm LTE}$, excitation temperature and optical depth of CO molecular lines by LTE analysis. The excitation temperature and column density distribution from LTE analysis are similar to the dust temperature and column density distribution derived from Herschel data. 

3. {\it Dendrogram} algorithm was conducted to extract the density structures at different scales (leaves and branches) from CO molecular lines in PPV space. The virial parameter of {\it Dendrogram} structures show most of structures are gravitationally bounded, indicating the dominance of gravity in W33-blue. 
% PDFs of N$_{\rm LTE}$ and N$_{\rm PPMAP}$ for the entire region and six sub-regions also support the dominance of gravity. 
We discuss three kinds of velocity dispersion $\sigma_{\rm M2}$ extracted from moment-2 map, $\sigma_{\rm BTS}$ extracted from velocity dispersion cube output by BTS algorithm and v$_{\rm rms}$ calculated by {\it Dendrogram} algorithm in detail, and suggest that they may reflect different physics. $\sigma_{\rm M2}$ and N$_{\rm LTE}$ as homologous physical quantities show strong dependency, which supports the non-thermal motions originating from gravitationally driven collapse.

4. We identify the filament skeletons from the moment 0 maps of $^{13}$CO (1$-$0) by FILFINDER algorithm, and extract velocity and intensity from moment maps along the filaments. The velocity distribution extracted from moment-1 map can keep pace with the velocity field in PPV space.
The expected maximum velocity shear around the hub does not appear, which means the velocity gradients may be not dominated by the longitudinal inflow. 
After comparing the polycyclic aromatic hydrocarbons and molecular lines emission, we support that the projection isn't the only reason to cause large-scale velocity gradient in W33-blue. The velocity gradient may mainly originate from the velocity difference in W33-blue, which result from the cloud-cloud collision and feedback of active star formation.

5. We argue that the cloud-cloud collision may trigger the formation of hub-filament structure in W33 complex. W33-blue is still a very young star cluster and likely in the transition from the compressed layer to a hub-filament system, which is consistent with the ratio of the line mass to a critical line mass $f \sim 1$. Local V-shape velocity structures around local intensity peaks may reveal the existence of local collapse centers. The scenario of multiple-scale hub-filament systems indicates a self-organization process from small-scale hub-filament structures (such as SDC 13) to the large-scale one (W33-blue). 

\section*{Acknowledgements}
H.-L. Liu is supported by National Natural Science Foundation of China (NSFC) through the grant No.12103045. 
This publication makes use of data from FUGIN, FOREST Unbiased Galactic plane Imaging survey with the Nobeyama 45-m telescope, a legacy project in the Nobeyama 45-m radio telescope.

This publication is based on data acquired with the Atacama Pathfinder Experiment (APEX) under programmes 092.F-9315 and 193.C-0584. APEX is a collaboration among the Max-Planck-Institut fur Radioastronomie, the European Southern Observatory, and the Onsala Space Observatory. The processed data products are available from the SEDIGISM survey database located at https://sedigism.mpifr-bonn.mpg.de/index.html, which was constructed by James Urquhart and hosted by the Max Planck Institute for Radio Astronomy.

% This paper has benefited from helpful discussions with Enrique Vázquez-Semadeni. 
Thanks Erik Rosolowsky for helping me to understand the {\it Dendrogram} algorithm in details.

\section{Data availability}
The data underlying this article are available in the article and the public data release of SEDIGISM survey \citet{Schuller2021-500} and FUGIN survey \citet{Umemoto2017-69}.

\bibliography{ref}{}
\bibliographystyle{aasjournal}

% \begin{multicols}
\appendix

\section{Difference of $\sigma_{\rm M2}$ and  $\sigma_{\rm BTS}$}\label{app-A}
\begin{figure}  \includegraphics[width=0.47\textwidth]{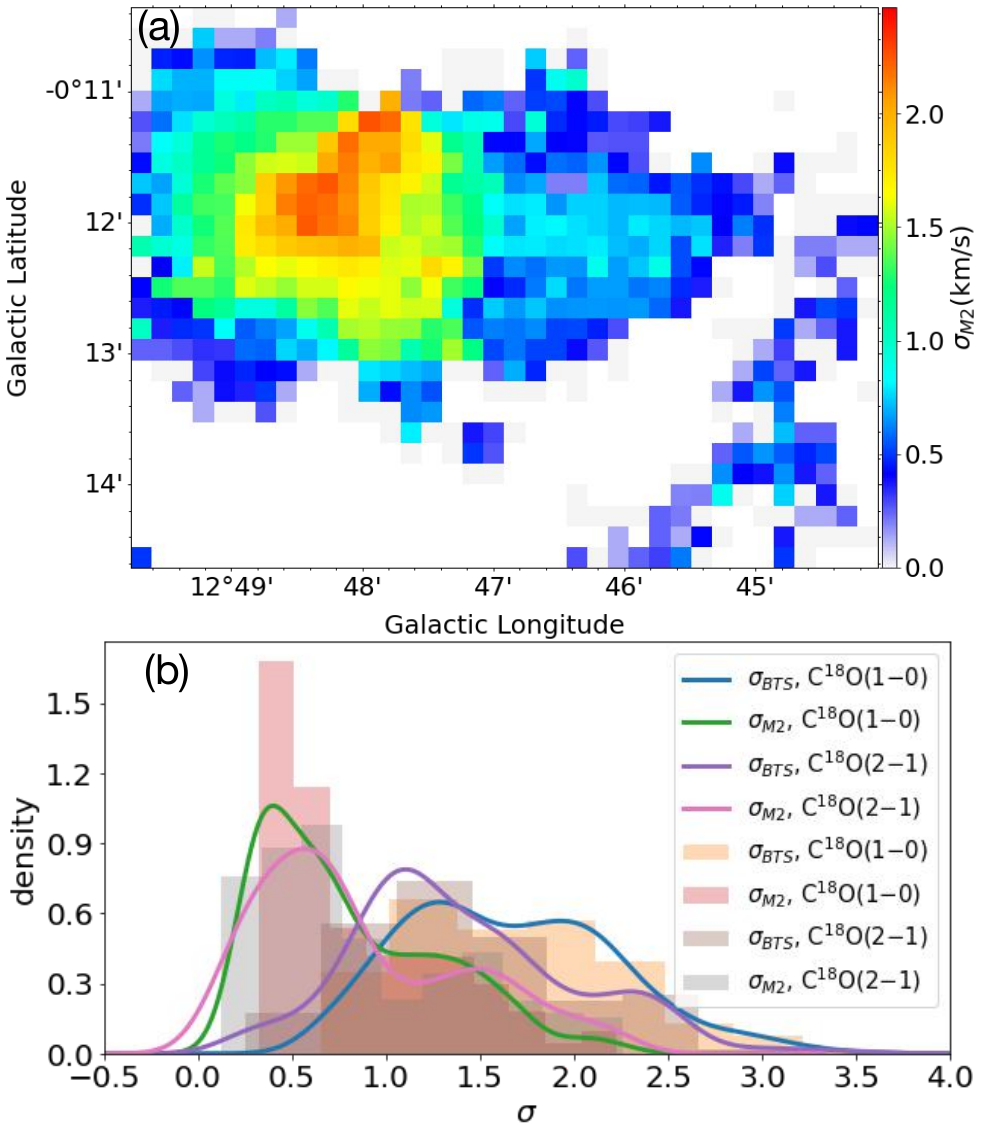}
\caption{(a) Velocity dispersion map of C$^{18}$O (2$-$1); (b) Distributions of $\sigma_{\rm M2}$ and  $\sigma_{\rm BTS}$ for C$^{18}$O (1$-$0) and C$^{18}$O (2$-$1) emission.}
\label{pm}
\end{figure}
\begin{figure*}
  \includegraphics[width=0.85\textwidth]{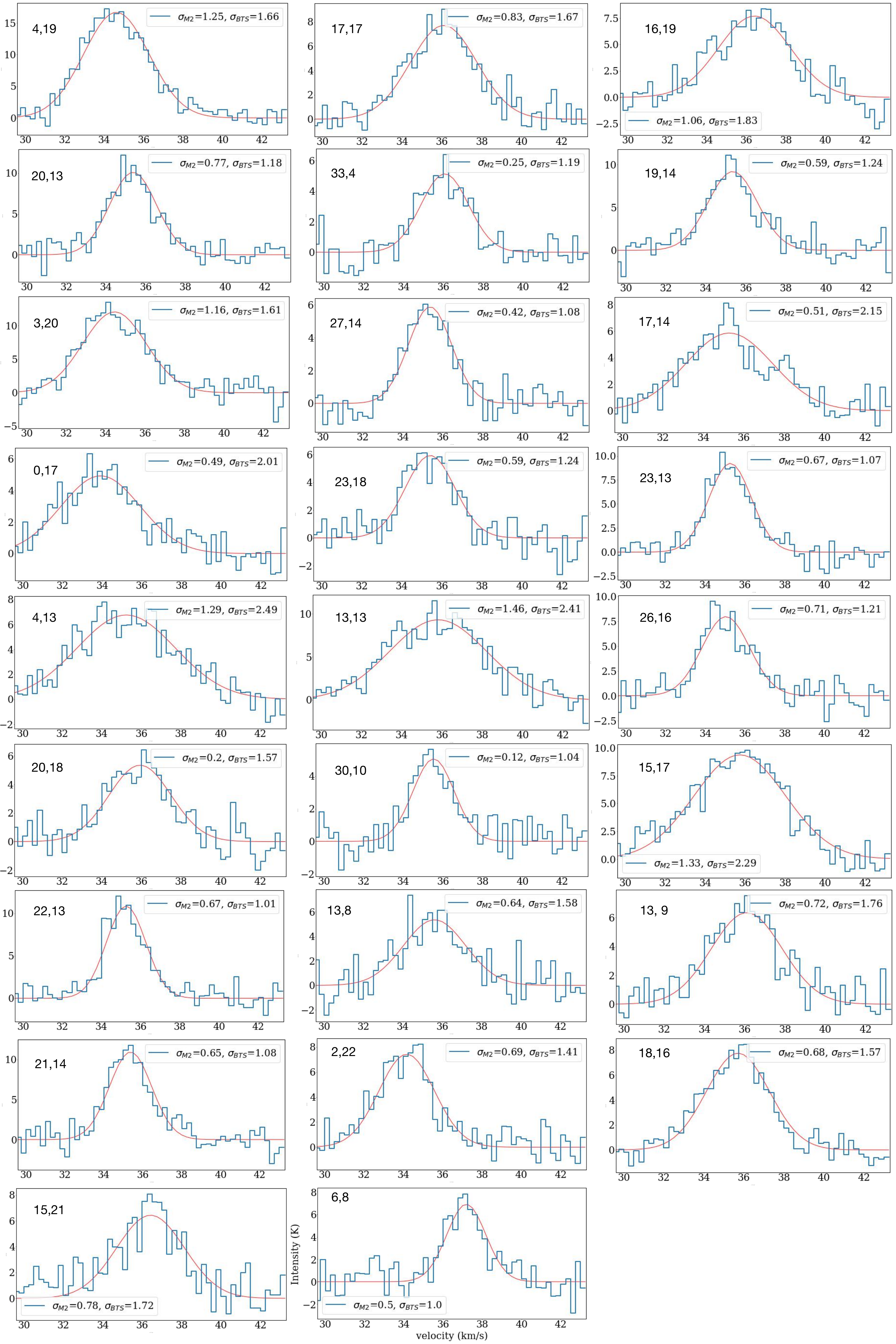}
\caption{Gaussian fitting of the C$^{18}$O (2$-$1) spectra at some pixels, upper left show the coordinates of pixels.}
\label{pf}
\end{figure*}

% For $^{18}$CO (1$-$0) emission, 
% the proportion of single Gaussian spectra is 93\%. Then we can shrink 3D velocity dispersion cube (output by BTS fitting) to 2D velocity dispersion map similar to moment-2 map by adding all velocity channels together. 
% more spectra are identified as double Gaussian, the proportion of single Gaussian spectra is 81\%. 

We cut out a sub-regions from C$^{18}$O (1$-$0) cube, and calculate two kinds of velocity dispersion $\sigma_{\rm M2}$ and  $\sigma_{\rm BTS}$ pixel-by-pixel. For excluding the influence of velocity resolution, we also compare C$^{18}$O (1$-$0) emission from FUGIN survey with C$^{18}$O (2$-$1) emission from SEDIGISM survey. 
At each pixel, we extract $\sigma_{\rm M2}$ and $\sigma_{\rm BTS}$ from 2D moment-2 map and 3D velocity dispersion cube output by BTS fitting, respectively. Here, $\sigma_{\rm BTS}$ won't have the issue of potential multiple velocity components. However, as shown in Fig.\ref{pm}(b), $\sigma_{\rm BTS}$ is still significantly larger than $\sigma_{\rm M2}$, 
consistent with the results in Sec.\ref{key}. Fig.\ref{pf} displays some specific examples of BTS fitting at some pixels, for all of them, Gaussian fitting gives significantly larger velocity dispersion than moment-2 map. This result does not rely on molecular line species and velocity resolution, because 4 molecular lines with different velocity resolution have been compared in this paper, and all of them show similar results. 

Actual spectral lines won’t have perfect Gaussian profiles, if we insist fitting them by Gaussian, generally will output larger velocity dispersion. Because one non perfect Gaussian profile can be decomposed into several perfect Gaussian profiles. However, we can’t decompose spectral lines endlessly in the actual situation, such as the limitation of velocity resolution. Though, the main reason may not be here. The difference of $\sigma_{\rm M2}$ and  $\sigma_{\rm BTS}$ may mainly come from the calculation methods. Both $\sigma_{\rm M2}$ and $\sigma_{\rm BTS}$ are calculated on the pixel scale, BTS algorithm is pixel-by-pixel fitting the spectra, while Moment-2 map is calculated for each data point directly. Moreover, moment map is intensity-weighted \footnote{\url{https://spectral-cube.readthedocs.io/en/latest/moments.html\#moment-maps}}, this aspect is also different from Gaussian fitting. The significant difference of velocity dispersion suggests these two methods are not equivalent, but it’s hard to say which one is better. As discussed in Sec.\ref{key}, velocity dispersion from these two methods may reflect different physics, rather than simply the pros or cons or alternative relationships. 

\section{Supplementary maps}
\begin{figure*}
  \includegraphics[width=0.85\textwidth]{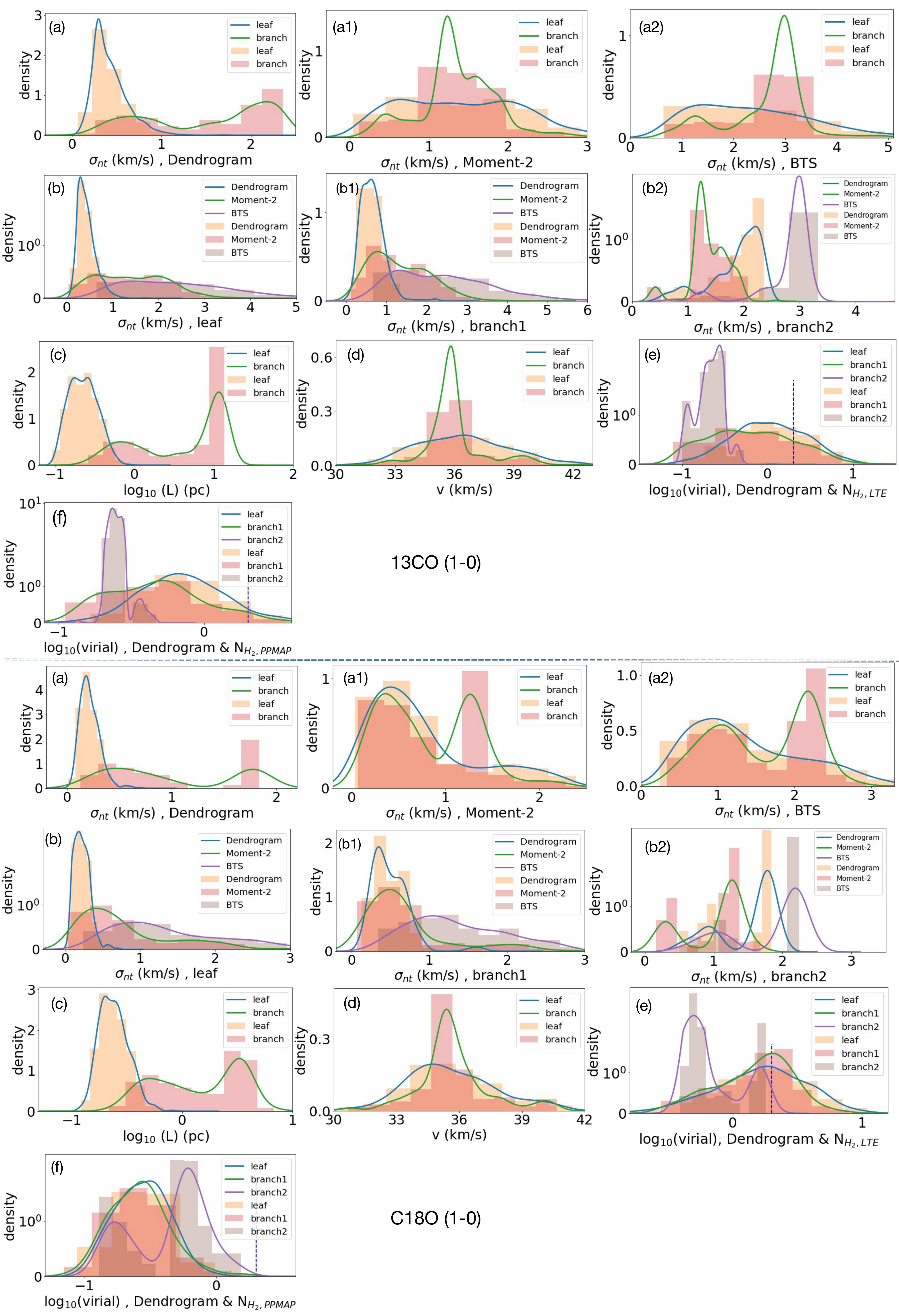}
\caption{(a) The non-thermal velocity dispersion v$_{\rm rms}$, $\sigma_{\rm BTS}$ and $\sigma_{\rm M2}$ of leaf and branch structures; (b) The comparison of non-thermal velocity dispersion v$_{\rm rms}$, $\sigma_{\rm BTS}$ and $\sigma_{\rm M2}$ for leaf and branch structures; (c) and (d) The size and mean velocity distribution of {\it Dendrogram} structures; (e) Virial parameters of {\it Dendrogram} structures calculated by v$_{\rm rms}$ and N$_{\rm LTE}$; (f) Virial parameters calculated by v$_{\rm rms}$ and N$_{\rm PPMAP}$. Dashed vertical blue line marks the position of the virial parameter $\alpha$=2.}
\label{co}
\end{figure*}
\begin{figure*}
  \includegraphics[width=0.9\textwidth]{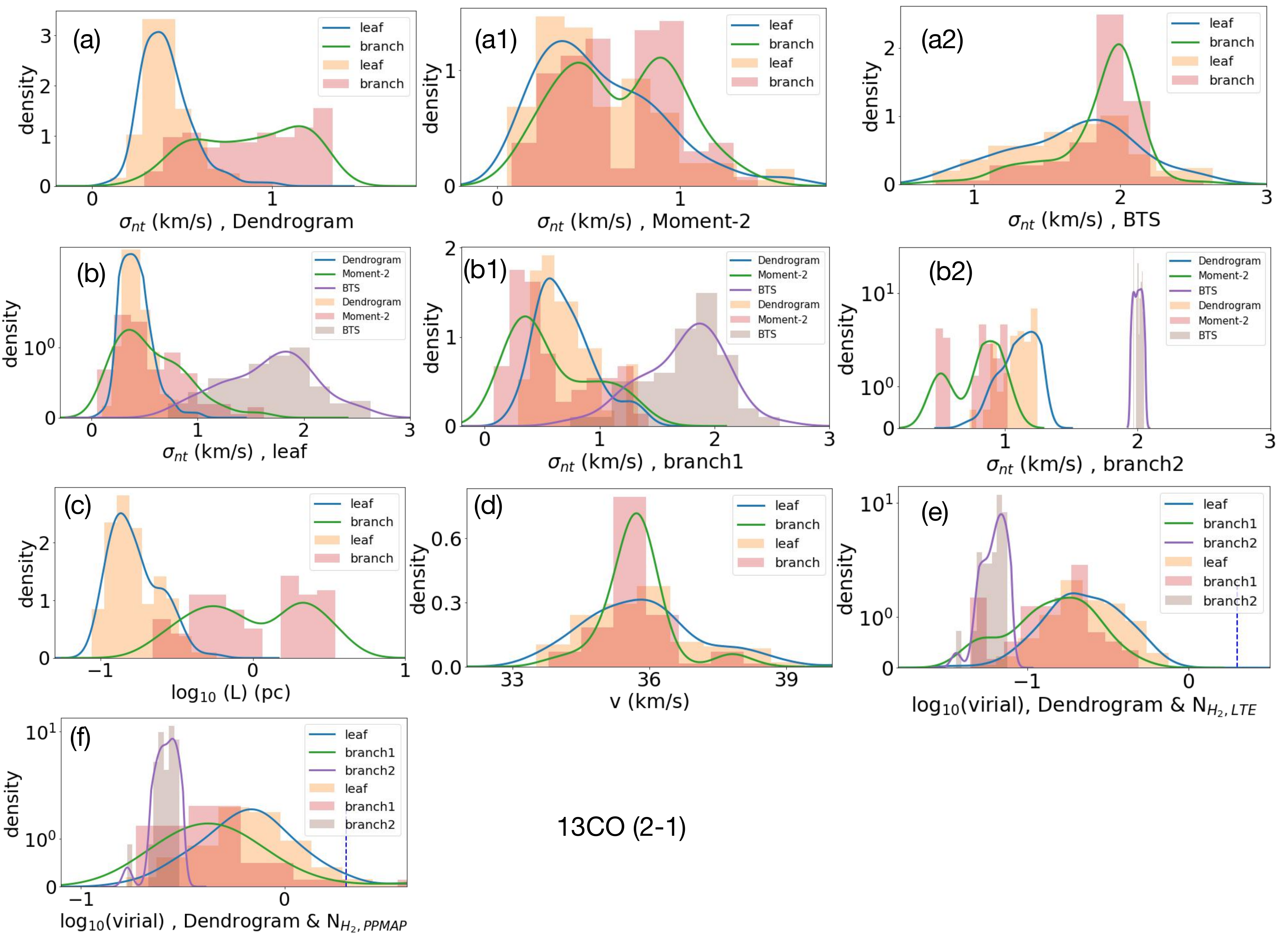}
\caption{The same with Fig.\ref{co}.}
% \label{co}
\end{figure*}

\begin{figure*}\includegraphics[width=1\textwidth]{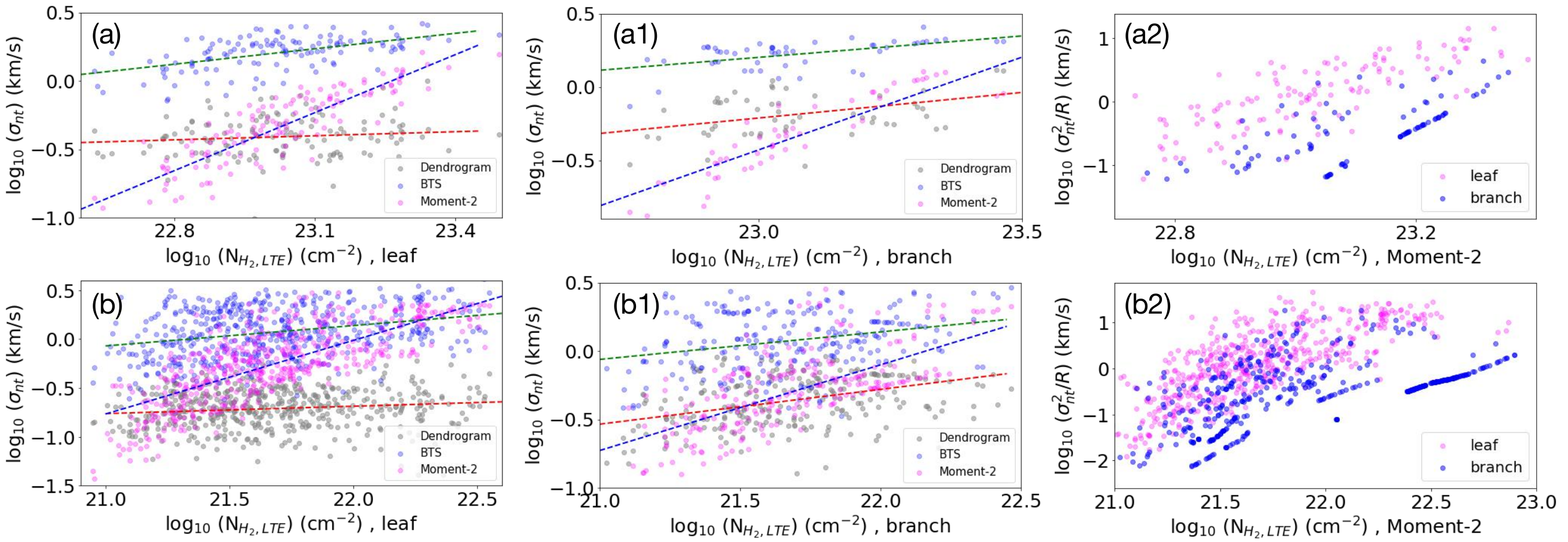}
\caption{ (a) and (b) are for C$^{18}$O (1-0) and $^{13}$CO (2-1), respectively.
First column: $\sigma_{nt}$--$N_{\rm H_2, LTE}$ relation of leaf structures. Gray, blue and magenta points represent v$_{\rm rms}$, $\sigma_{\rm BTS}$ and $\sigma_{\rm M2}$, respectively; Second column:  $\sigma_{nt}$--$N_{\rm H_2, LTE}$ relation of branch structures; Third column:  $\sigma_{nt}^2/R$--$N_{\rm H_2, LTE}$ relation, i.e. $a_{\mathrm{G}}$-- $a_{k}$ relation, here $\sigma_{nt}$ is estimated from $\sigma_{\rm M2}$. Blue and magenta points represent the branch and leaf structures.}
\label{sig-nn}
\end{figure*}

\begin{figure*}
 \includegraphics[width=1\textwidth]{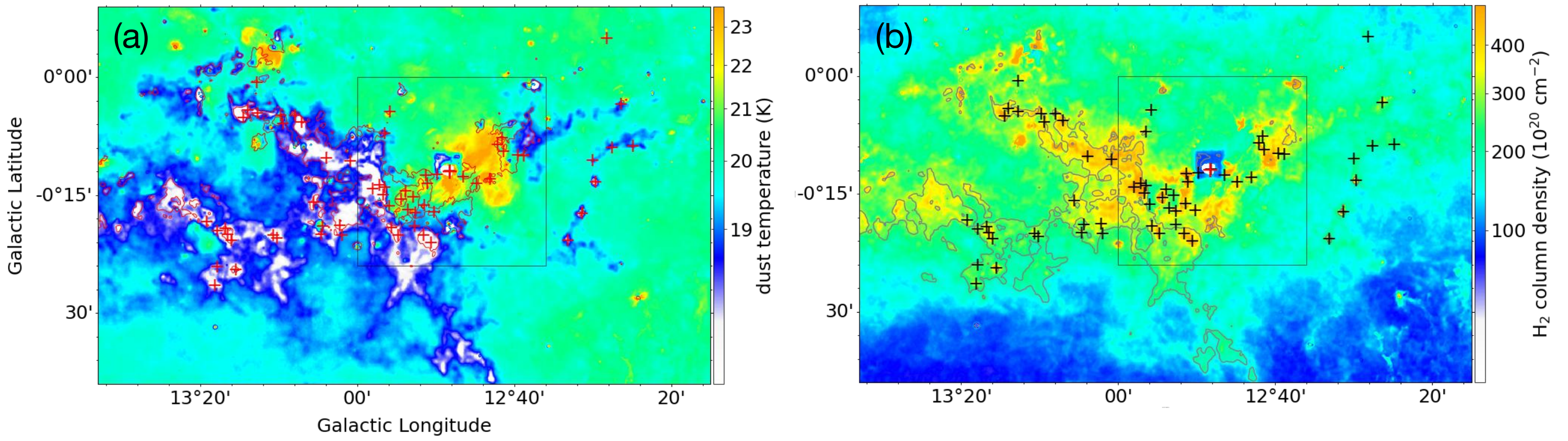}
\caption{Column density and temperature distribution in W33 complex derived from the PPMAP procedure. Boxes show the region investigated by \citet{Kohno2018-70S}, “+” marks the ATLASGAL clumps. The contour in (a) represents the column density with a level of 3.35$\times$10$^{22}$ cm$^{-2}$, the contour in (a) shows the dust temperature with a level of 18.6 K, in consistent with \citet{Dewangan2020-496}.}
\label{t-n}
\end{figure*}

\begin{figure*}
  \includegraphics[width=0.9\textwidth]{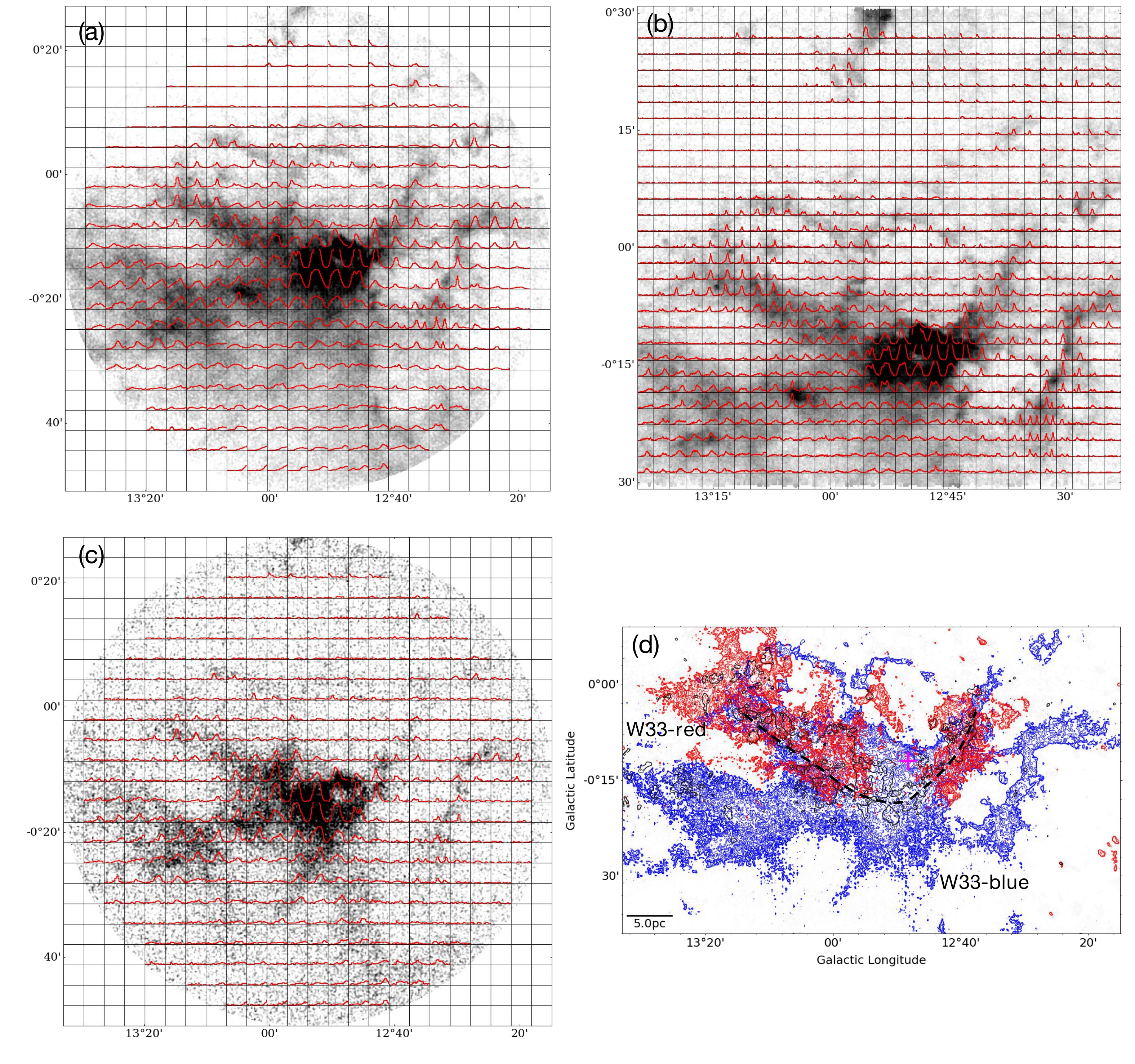}
\caption{ (a), (b) and (c) Grid maps of $^{13}$CO (1$-$0), C$^{18}$O (1$-$0) and $^{13}$CO (2$-$1). Background is the integrated intensity map; (d) Contours of $^{13}$CO (1$-$0) integrated intensity show two colliding clouds. The integrated velocity ranges for  blue and red contours are [29.6, 43.3] km/s and [47.2, 60.2] km/s, in consistent with \citet{Dewangan2020-496}. 
Blue and red contours represent W33-blue and W33-red marked in Fig.\ref{vf}(a) and Fig.\ref{bts}(a), respectively. Dashed line and black contours are the same with Fig.\ref{vf}(a).} 
\label{grid}
\end{figure*}

% \begin{figure*}
%   \includegraphics[width=0.98\textwidth]{w33/tree.pdf}
% \caption{(a) Dendrogram of W33-blue on $^{13}$CO(J =1$-$0) spectral cube; (b) The seg4 marked in (a).}
% \label{tree}
% \end{figure*}
% \begin{figure*}
%   \includegraphics[width=0.98\textwidth]{w33/leaf-branch.pdf}
% \label{leaf-branch}
% \end{figure*}

% \section{Column density}\label{lte-e}

\begin{figure*}
%   \centering
  \includegraphics[width=1\textwidth]{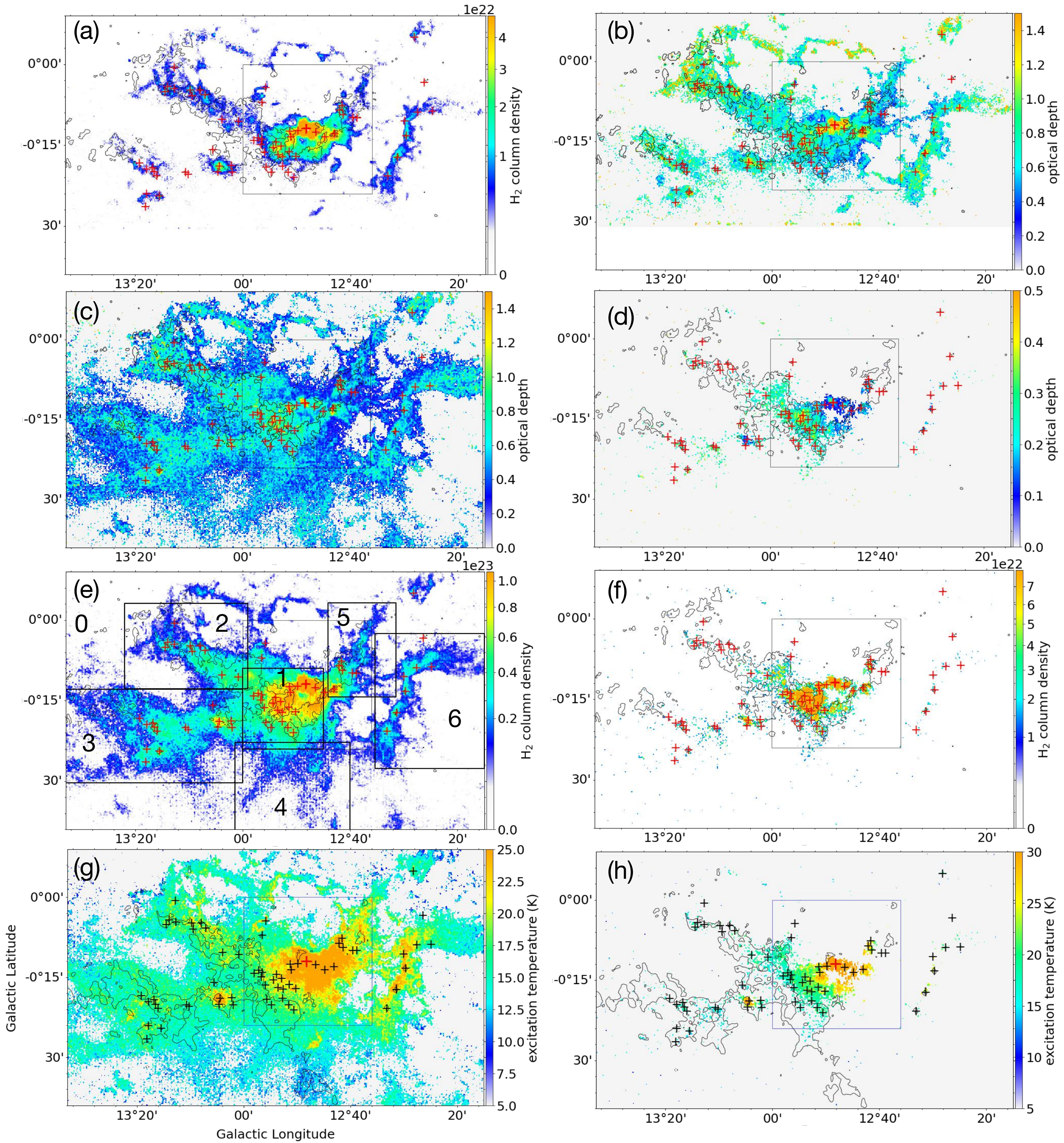}
\caption{(a) and (b): the H$_{2}$ column density and optical depth distributions in W33-blue derived from the $^{13}$CO (2$-$1) emission by LTE analysis; The remaining three lines show the optical depth, H$_{2}$ column density and excitation temperature distributions derived from $^{13}$CO (1$-$0) and C$^{18}$O (1$-$0) emission. The abundance ratio X$_{\rm ^{13}CO}$ and X$_{\rm C^{18}O}$ of H$_2$ compared with $^{13}$CO and C$^{18}$O are $\sim 7.1 \times 10^5$ and $\sim 5.9 \times 10^6$ respectively \citep{Frerking1982-262}.
Box shows the region investigated by \citet{Kohno2018-70S}, “+” marks the ATLASGAL clumps. Sub-regions marked in (e) are the same with Fig.\ref{vf}(d).}
\label{lte}
\end{figure*}
% \section{Velocity gradient}

\begin{figure*}
\includegraphics[width=0.8\textwidth]{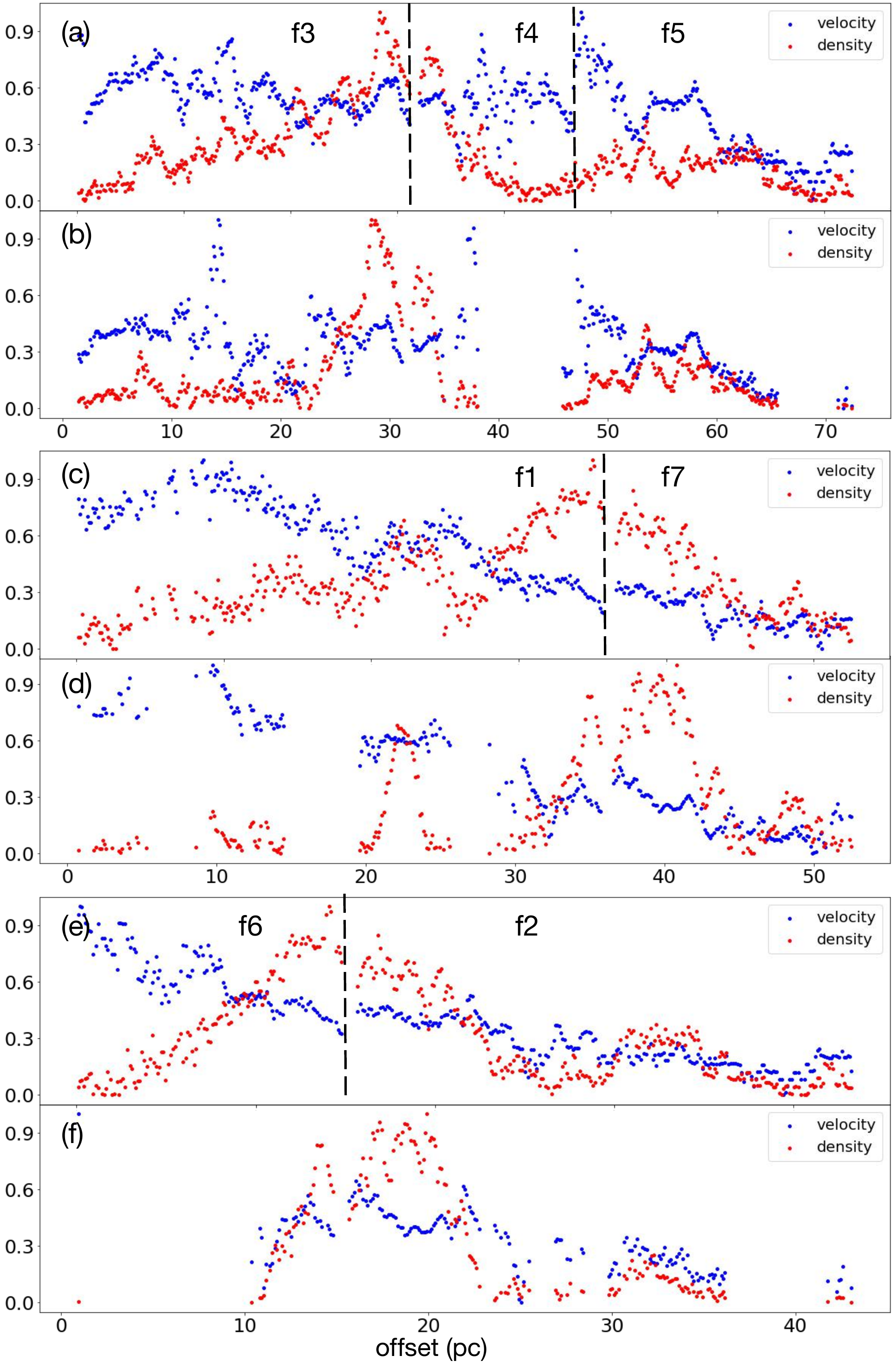}
\caption{(a), (c) and (e): distributions of  normalized velocity and intensity extracted from the moment 0 and moment 1 maps of $^{13}$CO (1$-$0) emission along the filament skeletons marked in Fig.\ref{vf}; (b), (d) and (f): distributions of  normalized velocity and intensity extracted from the moment 0 and moment 1 maps of $^{13}$CO (2$-$1) emission along the filament skeletons.}
\label{vg}
\end{figure*}

\clearpage

\noindent
Author affiliations:\\
$^{1}$National Astronomical Observatories, Chinese Academy of Sciences, Beijing 100101, Peoples Republic of China \\
$^{2}$University of Chinese Academy of Sciences, Beijing 100049, Peoples Republic of China  \\
$^{3}$Max Planck Institute for Astronomy, K\"onigstuhl 17, 69117 Heidelberg, Germany\\
$^{4}$Department of Astronomy, Yunnan University, Kunming, 650091, PR China\\
$^{5}$ Department of Physics, Faculty of Science, Kunming University of Science and Technology, Kunming 650500, People's Republic of China\\
$^{6}$Kavli Institute for Astronomy and Astrophysics, Peking University, Haidian District, Beijing 100871, People’s Republic of China\\
$^{7}$Department of Astronomy, School of Physics, Peking University, Beijing 100871, People’s Republic of China\\
$^{8}$Institute of Astronomy and Astrophysics, Anqing Normal University, Anqing, 246133, PR China\\
$^{9}$Shanghai Astronomical Observatory, Chinese Academy of Sciences, 80 Nandan Road, Shanghai 200030, Peoples Republic of China

\bsp	% typesetting comment
\label{lastpage}
\end{document}